\documentclass[fleqn,usenatbib]{mnras}

\usepackage{newtxtext,newtxmath}

\usepackage[T1]{fontenc}

\DeclareRobustCommand{\VAN}[3]{#2}
\let\VANthebibliography\thebibliography
\def\thebibliography{\DeclareRobustCommand{\VAN}[3]{##3}\VANthebibliography}

\usepackage{graphicx}	
\usepackage{amsmath}	
\usepackage{makecell}
\usepackage{multicol}
\usepackage{multirow}
\usepackage{float}
\usepackage{stfloats}
\usepackage[normalem]{ulem}

\defcitealias{sun_evolution_2024}{S24}
\defcitealias{davies_jwst_2024}{D24}

\title[Relic outflows in $1.5 < z < 5$ quiescent galaxies]{The \textit{JWST} EXCELS survey: Outflows in $1.5 < z < 5$ quiescent and recently quenched galaxies are likely relics from episodic AGN activity}

\author[E. Taylor et al.]{Elizabeth Taylor,$^{1}$ Adam C. Carnall,$^{1}$ David Maltby,$^{2}$ Omar Almaini,$^{2}$ Ho-Hin Leung,$^{1}$\newauthor Struan D. Stevenson,$^{1}$ Andrea Negri,$^{3}$ Fergus Cullen,$^{1}$ Vivienne Wild,$^{4}$ Ross J. McLure,$^{1}$\newauthor Alice E. Shapley,$^{5}$ Karla Z. Arellano-Córdova,$^{1}$ Ryan Begley,$^{6}$ Cecilia Bondestam,$^{1}$ Thomas de Lisle,$^{2}$\newauthor Callum T. Donnan,$^{7}$ James S. Dunlop,$^{1}$ Richard Ellis,$^{8}$ Guillaume Hewitt,$^{2}$ Anton M. Koekemoer,$^{9}$\newauthor Feng-Yuan Frey Liu,$^{1}$ Derek J. McLeod,$^{1}$ Kate Rowlands,$^{10, \,11}$ Ryan L. Sanders,$^{12}$ Dirk Scholte,$^{1}$\newauthor Maya Skarbinski,$^{11}$  and Thomas M. Stanton$^{1}$
\\
$^{1}$Institute for Astronomy, University of Edinburgh, Royal Observatory, Edinburgh, EH9 3HJ, UK\\
$^{2}$University of Nottingham, School of Physics \& Astronomy, Nottingham, NG7 2RD, UK \\
$^{3}$Facultad de F\'isicas, Universidad de Sevilla, Avda. Reina Mercedes S/N, Campus de Reina Mercedes, E-41012, Seville, Spain \\
$^{4}$School of Physics \& Astronomy, University of St Andrews, North Haugh, St Andrews, KY16 9SS, UK \\
$^{5}$Department of Physics \& Astronomy, University of California, Los Angeles, 430 Portola Plaza, Los Angeles, CA 90095, USA \\
$^{6}$Armagh Observatory and Planetarium, College Hill, Armagh, BT61 9DG, N. Ireland, UK \\
$^{7}$NSF’s National Optical-Infrared Astronomy Research Laboratory, 950 N. Cherry Ave., Tucson, AZ 85719, USA \\
$^{8}$University College London, Department of Physics \& Astronomy, Gower Street, London WC1E 6BT, UK \\
$^{9}$Space Telescope Science Institute, 3700 San Martin Drive, Baltimore, MD 21218, USA \\
$^{10}$AURA for ESA, Space Telescope Science Institute, 3700 San Martin Drive, Baltimore, MD 21218, USA \\
$^{11}$William H. Miller III Department of Physics and Astronomy, Johns Hopkins University, Baltimore, MD 21218, USA \\
$^{12}$Department of Physics and Astronomy, University of Kentucky, 505 Rose Street, Lexington, KY 40506, USA
}

\date{Accepted XXX. Received YYY; in original form ZZZ}

\pubyear{\the\year{2025}}

\begin{document}
\label{firstpage}
\pagerange{\pageref{firstpage}--\pageref{lastpage}}
\maketitle

\begin{abstract}
We investigate the presence and origin of neutral gas outflows and inflows in 13 post-starburst (PSB) and quiescent galaxies at redshifts $1.8 \leq z \leq 4.6$, using \textit{JWST} NIRSpec spectroscopy from the EXCELS survey. Na$\,$\textsc{d} absorption profiles reveal that 3 out of 13 exhibit blueshifted absorption indicative of outflows, and a further 2 objects show signs of inflowing gas. Outflow velocities range from $\approx$ 300\,--\,1200\,km\,s$^{-1}$, and we find gas flows are detected exclusively in objects that quenched $< 600$\, Myr ago. We derive mass outflow rates over two orders of magnitude higher than current levels of star formation in our sample, indicating that the winds are unlikely to be driven by supernovae. The majority of the outflow sample have anomalously high energy and momentum outflow rates compared to those predicted for current levels of star formation or AGN activity. We conclude that we are likely observing fossil outflows driven by previous, more luminous AGN activity which has since faded. We then compare with the \textsc{eagle} simulation to explore a potential `outflow cycle', finding that our observations are consistent with a model in which $z\sim3$ quiescent galaxies undergo short $\simeq5$\,Myr periods of AGN activity strong enough to drive outflows, which occur every $\simeq$ 40\,Myr on average. This AGN activity drives observable outflows that persist for up to $\simeq 10$\,Myr after the AGN fades, followed by a $\simeq20$\,Myr lull, and a subsequent short inflow, which eventually re-ignites AGN activity, and the cycle repeats.
\end{abstract}




\section{Introduction}\label{section:intro}

The driving force behind the cessation of star-formation in galaxies, or `quenching', remains a key area of interest in astrophysics. Feedback from both active galactic nuclei (AGN) and intense star-formation are key ingredients in cosmological simulations of galaxy evolution and quenching \citep[e.g.,][]{springel_cosmological_2003, di_matteo_energy_2005, hopkins_unified_2006, schaye_eagle_2015, dave_simba_2019}, where it is invoked to prevent the build-up of too many massive galaxies. Generally, feedback is thought to clear cold gas from star-forming regions, and prevent further star formation once a galaxy is quenched, through the heating of gas via shocks and radiation \citep[see, e.g.,][]{silk_quasars_1998, fabian_observational_2012, sokal_prevalence_2016, alatalo_shocked_2016, harrison_observational_2024}. 

Despite the reliance of cosmological simulations on feedback prescriptions to produce realistic looking galaxy populations, observational studies of feedback are quite challenging \citep[for a recent review, see][]{veilleux_cool_2020}. Galactic-scale outflowing winds, detected through analysis of emission and absorption lines in spectroscopic data, are a signature of feedback processes. Ionized gas outflows are commonly traced through forbidden emission lines such as [O$\,$\textsc{iii}] and recombination lines (e.g., H$\alpha$), but winds in this phase alone are unlikely to fully quench a galaxy \citep[e.g.,][]{carniani_fast_2016, leung_mosdef_2019, lamperti_super_2021, concas_being_2022}. Cool neutral and molecular outflows, tracing gas at temperatures $T \lesssim 10^{4}$\,K via UV/optical absorption lines (e.g., Na$\,$\textsc{d}, Mg$\,$\textsc{ii}, Fe$\,$\textsc{ii}), molecules (e.g., CO), and H\,\textsc{i} are thought to dominate the mass and energetics of galactic winds \citep[e.g.,][]{vayner_galactic-scale_2017, herrera-camus_molecular_2019, kim_first_2020}. While outflows are multiphase and complex in nature, the majority of studies focus on single gas phases, largely due to limited access to multiple tracers.

Locally, studies of outflowing neutral gas exploit the Na$\,$\textsc{d} ($\lambdaup\lambdaup$\,5891,\,5897\,\AA) absorption profile \citep[e.g.,][]{forster_echelle_1995, rupke_outflows_2005, chen_absorption-line_2010, concas_two-faces_2019, perna_muse_2020, fluetsch_properties_2021, baron_multiphase_2022, sun_evolution_2024}, however this feature is shifted into the NIR at higher redshifts. Some works have also utilised detections of the 21\,cm H\,\textsc{i} line in absorption to trace neutral gas outflows, however such detections are challenging \citep[see, e.g.,][for a detailed review]{morganti_interstellar_2018}. At 0.5 $\lesssim z \lesssim $ 1.5, outflow studies have typically been performed using rest-frame UV spectral features shifted into the optical bands (e.g., Mg$\,$\textsc{ii}). Spectral stacking of large numbers of galaxies is often undertaken due to lower signal-to-noise from ground-based surveys, for both ionized and cool neutral tracers \citep[][]{talia_gmass_2012,bradshaw_high-velocity_2013, cicone_outflows_2016, belli_mosfire_2019, prusinski_connecting_2021}.

At cosmic noon, and in the more local Universe, extremely high-velocity outflows ($\Delta v > 1000$\,km\,s$^{-1}$) have been seen in star-forming galaxies both with AGN \citep[e.g.,][]{harrison_energetic_2012, cicone_massive_2014, talia_agn-enhanced_2017} and without \citep{bradshaw_high-velocity_2013, geach_stellar_2014}. Outflows within galaxies that have very recently stopped forming stars, such as post-starbursts (PSBs), may provide insights into whether such winds are capable of quenching a galaxy directly. PSBs are typically identified by their spectra showing both strong Balmer absorption lines typical of a population dominated by young A/F stars, and weak or absent nebular emission lines, indicating the lack of ongoing star formation \citep[see, e.g.,][]{dressler_spectroscopy_1983, wild_evolution_2016, french_evolution_2021}. High-velocity winds are detected in PSBs at cosmic noon \citep[e.g.,][]{ maltby_high-velocity_2019, taylor_high-velocity_2024}, $z \sim 0.7$ \citep[e.g.,][]{tremonti_discovery_2007}, and locally \citep[e.g.,][]{baron_multiphase_2022}. 

Some studies have used PSBs to attempt to pin down how outflows may affect the evolution of a galaxy, and whether these winds are directly linked to the primary quenching event. In the local Universe, \cite{harvey_cool_2025} found excess cool gas absorption up to 1\,Mpc from massive PSBs, and suggest that the high-velocity winds observed in these galaxies may even impact the circum-galactic medium. \citeauthor{sun_evolution_2024} (\citeyear{sun_evolution_2024}, hereinafter \citetalias{sun_evolution_2024}) used a sample of galaxies from the Sloan Digital Sky Survey (SDSS) to form an evolutionary sequence of declining star formation rate, selecting PSBs and their likely progenitors and descendants based on star-formation rates and burst mass fractions. The authors found outflow velocities decrease with time elapsed since the end of a starburst.

Conversely, \cite{taylor_high-velocity_2024} found high-velocity winds ($\Delta v > 1000$\,km\,s$^{-1}$) persisted in PSBs at $z > 1$ up to 1 Gyr after the last burst of star formation. An outflow travelling at $\simeq1000$ km s$^{-1}$ will cross the $\simeq1$ kpc extent of a high-redshift quiescent galaxy in just 1 Myr, suggesting it is unlikely these outflows were launched at the time of quenching. It should be noted that \cite{taylor_high-velocity_2024} investigated outflows traced by Mg$\,$\textsc{ii}, while \citetalias{sun_evolution_2024} use Na$\,$\textsc{d}, and the choice of outflow tracer may affect results \citep[e.g.,][]{kornei_properties_2012}. Despite the difference in findings, both studies attribute these winds to AGN. \citetalias{sun_evolution_2024} argue that, since star-formation rate (SFR) in their sample declines faster than outflow velocity, the outflows are likely driven by AGN, while \cite{taylor_high-velocity_2024} conclude that their observations are of relic winds driven by past episodic AGN activity. A consideration of the energetics suggests the presence of high-velocity outflows in PSBs is not necessarily inconsistent with the low levels of AGN activity observed, if the duty cycle for powerful AGN activity is short \citep[][]{almaini_no_2025}. 

While PSBs are rare at low redshift (comprising $\lesssim 1$ per cent of local massive galaxies), they become more prevalent as a fraction of the overall passive population at $z\gtrsim1$ \citep[$\approx5$ per cent; e.g.,][]{wild_evolution_2016, taylor_role_2023}. At higher redshift ($z >$ 3), the launch of \textit{JWST} has indeed revealed an unexpected abundance of recently quenched and quiescent galaxies, both through deep imaging surveys \citep[e.g.,][]{carnall_surprising_2023, alberts_high_2024, long_efficient_2024, stevenson_primer_2025} and spectroscopy \citep[e.g.,][]{carnall_massive_2023, carnall_jwst_2024, glazebrook_massive_2024, nanayakkara_population_2024, setton_uncover_2024, weibel_rubies_2025, de_graaff_efficient_2025}, allowing in-depth studies of quenching mechanisms in galaxies at cosmic noon and beyond. Of particular note for studies of cool, neutral outflows, the wavelength coverage of \textit{JWST} has enabled the observation of Na$\,$\textsc{d} absorption lines out to $z \simeq 7-8$, whilst the increase in signal-to-noise allows the study of Mg$\,$\textsc{ii} in individual objects at $z > 1$, without the need for stacking \citep[e.g.,][]{wu_ejective_2025, valentino_gas_2025, kehoe_aurora_2025}.

Recent works using \textit{JWST} have made use of the Na$\,$\textsc{d} absorption profile to detect outflow signatures in mixed samples of star-forming and quiescent galaxies at $z >$ 2 \citep[e.g.,][hereinafter \citetalias{davies_jwst_2024}]{davies_jwst_2024}, with some studies focusing on individual post-starburst and recently quenched objects \citep[e.g.,][]{belli_star_2024, sun_extreme_2026, valentino_gas_2025, wu_ejective_2025}. The consensus on the driver behind outflows in these galaxies is, again, mixed, with some being attributed to current AGN feedback \citep{belli_star_2024, wu_ejective_2025}, and others to relic winds from star formation or episodic AGN activity \citep{valentino_gas_2025, sun_extreme_2026}. 

How galactic winds are related to the quenching of galaxies is still an open question. In this work, we utilise the power of \textit{JWST} in an attempt to constrain drivers of galactic-scale winds at high redshift ($1.5 < z < 5$). We use Na$\,$\textsc{d} absorption lines to study the incidence of outflows in a sample of 13 PSBs and quiescent galaxies from the \textit{JWST} EXCELS survey, along with comparable objects from recent literature. We consider the outflow energetics and relationship to potential indicators of AGN activity, as well as how these evolve with time since quenching took place.

The structure of this paper is as follows. In Section \ref{section:data} we present our data and sample selection, followed by our sodium absorption profile fitting methodology in Section \ref{section:analysis}. We present our results in Section \ref{section:results}, and discuss outflow energetics further in Section \ref{section:energetics}. In Section \ref{sec:time} we investigate possible outflow timescales, ending with our conclusions and a brief summary in Section \ref{section:conc}. Throughout this paper, we adopt the AB magnitude system \citep[][]{oke_secondary_1983} and a flat $\Lambda$CDM cosmology with $\Omega_M$ = 0.3, $\Omega_{\Lambda}$ = 0.7, and $H_0 = 100$\,\textit{h}\,km\,s$^{-1}$\,Mpc$^{-1}$, where $h$ = 0.7. We assume a \cite{kroupa_variation_2001} initial mass function, and assume a solar abundance of $\mathrm{Z}_{\odot} = 0.0142$ \citep[][]{asplund_chemical_2009}.

\section{Data and Sample Selection}\label{section:data}

\subsection{\textit{JWST} EXCELS}\label{sub:EXCELS}

The spectroscopy used in this work is drawn from the \textit{JWST} Early eXtragalactic Continuum and Emission Line Science (EXCELS; GO 3453; PIs: A. Carnall \& F. Cullen) survey \citep[][]{carnall_jwst_2024}. EXCELS is a NIRSpec Cycle 2 programme, designed as a follow-up to the \textit{JWST} Public Release IMaging for Extragalactic Research (PRIMER) Cycle 1 programme \citep{dunlop_primer_2021}, a multiband NIRCam survey targeting regions of the UDS and COSMOS CANDELS legacy fields \citep[][]{grogin_candels_2011, koekemoer_candels_2011}. 

EXCELS consists of four NIRSpec Multi-Shutter Assembly (MSA) pointings within the PRIMER-UDS field, observed with three combinations of gratings and filters (G140M/F100LP, G235M/F170LP and G395M/F290LP) at medium resolution (\textit{R} $\sim$ 1000). The total exposure times per pointing for the G140M and G395M gratings are $\sim$ 4 hours, and $\sim$ 5.5 hours for the G235M grating. Observations were carried out using a 3-shutter slitlet and 3-point nodding pattern. Full details of the survey selection and observation process are presented in \cite{carnall_jwst_2024}.

The EXCELS target selection was based on both the VANDELS UDS-\textit{HST} \citep{mclure_vandels_2018} and the PRIMER-UDS \citep[][]{dunlop_primer_2021, begley_evolution_2025} photometric catalogues. The two relevant EXCELS target classes for this paper are the VANDELS cosmic noon passive sample and the PRIMER ($z > 2$) passive sample. The VANDELS sample was selected by CANDELS photometric redshift and UVJ colour, using the \cite{williams_detection_2009} criteria. The PRIMER sample was selected by fitting star-formation history models to candidate photometry, applying a specific star-formation rate (sSFR $= \mathrm{SFR}/M_{*}$) cut of sSFR < 0.2 / $t_{\mathrm{H}}(z)$, where $t_{\mathrm{H}}(z)$ is the age of the Universe as a function of the observed redshift. We refer the reader to \cite{mclure_vandels_2018} and \cite{carnall_jwst_2024} for more details.  

Spectroscopic data used in this work are reduced using the methodology detailed in Leung et al. (in prep), extended to include two sources at $z < 3$. In brief, Leung et al. process the data products using the default JWST pipeline (v1.19.1, CRDS\_CTX = jwst\_1413.pmap) at all levels, with a custom \texttt{clean\_flicker\_noise} step at level 1, and additional automatic and manual pixel masking based on certain data quality bit-masks at level 2. A custom optimal extraction method \citep{horne_optimal_1986} with a wavelength-varying kernel is used to extract 1D spectra. Leung et al. combine separate gratings by first scaling the G140M and G395M spectra to the G235M spectrum. Higher-resolution gratings have their resolution degraded to match the lower-resolution gratings in overlap regions using \textsc{SpectRes} \citep{carnall_spectres_2017}, and the spectra are joined by averaging flux in overlapping regions. A best-fitting SED is then found for each object, through fitting the available aperture-corrected and point spread function (PSF) homogenised \textit{HST} and \textit{JWST} photometry using the \textsc{Bagpipes} fitting code \citep{carnall_inferring_2018, carnall_vandels_2019}. Each spectrum is scaled to the ratio of the piecewise median of the best-fitting model using a 15$^{\mathrm{th}}$--order Chebyshev polynomial.

\subsection{Sample Selection}\label{sub:sample}

\begin{figure}
    \centering
    \includegraphics[width=\columnwidth]{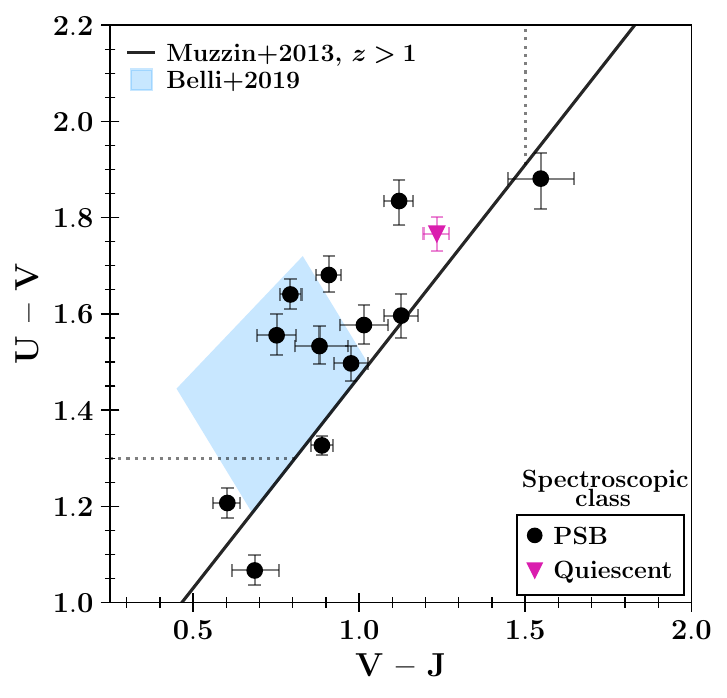}
    \caption{Rest-frame UVJ diagram for our sample. The $U - V$ and $V - J$ colours are calculated via star-formation history modelling using the \textsc{Bagpipes} code (see Section \ref{sub:properties}). PSBs selected with W$_{\textrm{H$\delta$}} > 5$\,\AA\ are depicted as black circles, and older quiescent galaxies selected with W$_{\textrm{H$\delta$}}$ < 5\,\AA\ and W$_{\textrm{[O$\,$\textsc{ii}]}} > -$5\,\AA\ are shown as magenta triangles. The solid black line depicts the quiescent selection criteria for galaxies with redshifts in the range 1 $< z <$ 4 from \protect\cite{muzzin_evolution_2013}. The dashed horizontal and vertical lines represent further selection criteria from \protect\cite{williams_detection_2009}. The shaded blue region represents a post-starburst selection box for galaxies with a median stellar age of 0.3 $< t_{50} <$ 1\,Gyr \protect\citep[][]{belli_mosfire_2019}.}
    \label{fig:uvj}
\end{figure}

The full spectroscopic sample analysed in this work is composed of PSBs and quiescent galaxies with older stellar populations (as indicated by their H$\delta$ equivalent width; see below), for which we follow two different selection procedures, as described below. Our initial sample from the EXCELS catalogue is selected requiring an EXCELS type of `passive' \citep[see Section \ref{sub:EXCELS};][]{carnall_jwst_2024}, identified in either VANDELS or PRIMER, with secure spectroscopic redshifts. We include one additional galaxy selected for observation in EXCELS based on the PCA selection method for post-starburst galaxies detailed in \cite{wild_evolution_2016}. This results in an initial parent sample of 38 galaxies. 

The H$\delta\, (\lambdaup$\,4102\,\AA) line is essential for our sample selection (see following text for details), thus we remove any galaxies without coverage of this feature. As this work investigates neutral outflows via the Na$\,$\textsc{d} absorption profile, we additionally remove objects without the Na$\,$\textsc{d} $\lambdaup\lambdaup$\,5891,\,5897\,\AA\ feature in the available spectra. We also remove any galaxies for which the Na$\,$\textsc{d} or H$\delta\, (\lambdaup$\,4102\,\AA) features fall within $\sim 50\,$\AA\ of a detector gap, or to either end of the spectrum. As not all EXCELS galaxies were observed using all three medium-resolution gratings, these cuts remove 22 objects in the following way:
\begin{itemize}
    \item 11 galaxies with no coverage of H$\delta$ (6 of which also had no Na$\,$\textsc{d} coverage),
    \item 7 further galaxies with coverage of H$\delta$, but no Na$\,$\textsc{d} coverage,
    \item 4 galaxies with Na$\,$\textsc{d} or H$\delta$ close to a detector gap/end of the spectrum.
\end{itemize}
This leaves a final parent sample of 16 galaxies. 

We then select our PSB sample using cuts on the equivalent width of the H$\delta\, (\lambdaup$\,4102\,\AA) line, W$_{\textrm{H$\delta$}}$. To measure W$_{\textrm{H$\delta$}}$, we follow the non-parametric method of \cite{goto_h-strong_2003}, by shifting the spectra to the rest-frame, and subsequently using:
\begin{equation}\label{eq:ew}
    \mathrm{W} = \int_{\lambda_{1}}^{\lambda_{2}}\, 1 - \frac{F(\lambda)}{F_{c}(\lambda)}\, d\lambda,
\end{equation}
where $F(\lambda)$ is the spectral flux, and $F_{c}(\lambda)$ is the continuum flux. For W$_{\textrm{H$\delta$}}$, $\lambda_{1}$ and $\lambda_2$ are 4082 and 4122\,\AA, respectively. $F_{c}(\lambda)$ is estimated using a linear interpolation across the line region, using continuum midpoints determined from adjacent wavelength intervals. In the case of W$_{\textrm{H$\delta$}}$ these are 4030\,--\,4082 and 4122\,--\,4170\,\AA, as these intervals are free from significant absorption and emission lines. In this work, a positive (negative) equivalent width value indicates absorption (emission). Uncertainties on W$_{\textrm{H$\delta$}}$ are calculated through bootstrapping: we perturb each data point by the associated uncertainties and recalculate W$_{\textrm{H$\delta$}}$ 100 times, to find the standard deviation. We select PSBs with the commonly used criterion of W$_{\textrm{H$\delta$}}$ > 5\,\AA\ \citep[e.g.,][]{balogh_differential_1999, goto_catalogue_2007, french_evolution_2021}, which indicates a substantial population of A type stars, and thus a burst of star formation in the past few hundred Myr. This results in a final PSB sample of 12 galaxies.

\begingroup
\renewcommand{\arraystretch}{1.5}
\begin{table*}
    \centering
    \begin{tabular}{lcccccccccccccc}
\hline
    \makecell{EXCELS--ID \\ } & & \makecell{$z_{\mathrm{spec}}$ \\ } & & \makecell{W$_{\textrm{H$\delta$}}$ \\ (\AA)} & &
    \makecell{W$_{\textrm{[O$\,$\textsc{ii}]}}$ \\ (\AA)} & &
    \makecell{log$_{10}({M}_{*}/{\mathrm{M}}_{\odot})$ \\ } & & 
    \makecell{SFR \\ (M$_{\odot}$\,yr$^{-1}$)}  & &
    \makecell{$t_{\mathrm{sq}}$ \\ (Gyr)} & &
    \makecell{$r_{\mathrm{eff}}$ \\ (kpc)} \\
\hline
\hline

34495 & & 
3.797 & & 
\hphantom{0}9.26 $\pm$ 0.54 & & 
$-$4.02 $\pm$ 1.68 & & 
10.75$\genfrac{}{}{0pt}{}{+0.02}{-0.02}$ & & 
< 0.06\hphantom{\,\footnotesize{\textsuperscript{\textdagger}}} & & 
0.13$\genfrac{}{}{0pt}{}{+0.01}{-0.01}$ & & 
0.84 $\pm$ 0.02 \\

39063 & & 
3.703 & & 
\hphantom{0}9.63 $\pm$ 0.66 & & 
$-$0.91 $\pm$ 2.05 & & 
10.41$\genfrac{}{}{0pt}{}{+0.02}{-0.02}$ & & 
< 0.03\hphantom{\,\footnotesize{\textsuperscript{\textdagger}}} & & 
0.22$\genfrac{}{}{0pt}{}{+0.02}{-0.03}$ & & 
0.37 $\pm$ 0.01 \\

45981 & & 
3.083 & & 
\hphantom{0}6.41 $\pm$ 1.97 & & 
$-$1.79 $\pm$ 3.38 & & 
10.55$\genfrac{}{}{0pt}{}{+0.03}{-0.02}$ & & 
< 0.04\hphantom{\,\footnotesize{\textsuperscript{\textdagger}}} & & 
0.50$\genfrac{}{}{0pt}{}{+0.09}{-0.09}$ & & 
0.48 $\pm$ 0.01 \\

50789 & & 
3.990 & & 
10.92 $\pm$ 0.55 & & 
\hphantom{$-$}2.29 $\pm$ 2.34 & & 
10.98$\genfrac{}{}{0pt}{}{+0.03}{-0.02}$ & & 
< 0.10\hphantom{\,\footnotesize{\textsuperscript{\textdagger}}} & & 
0.15$\genfrac{}{}{0pt}{}{+0.02}{-0.02}$ & & 
1.13 $\pm$ 0.03 \\

55410\,$^{*}$ & & 
3.195 & & 
\hphantom{0}1.51 $\pm$ 1.22 & & 
$-$0.81 $\pm$ 1.78 & & 
11.04$\genfrac{}{}{0pt}{}{+0.02}{-0.02}$ & & 
< 0.11\hphantom{\,\footnotesize{\textsuperscript{\textdagger}}} & & 
0.50$\genfrac{}{}{0pt}{}{+0.03}{-0.03}$ & & 
1.35 $\pm$ 0.01 \\

57000 & & 
3.193 & & 
10.68 $\pm$ 1.63 & & 
$-$4.66 $\pm$ 3.63 & & 
10.76$\genfrac{}{}{0pt}{}{+0.04}{-0.03}$ & & 
< 0.36\,\footnotesize{\textsuperscript{\textdagger}} & & 
0.17$\genfrac{}{}{0pt}{}{+0.05}{-0.05}$ & & 
1.59 $\pm$ 0.01 \\

65915 & & 
4.362 & & 
\hphantom{0}9.04 $\pm$ 0.92 & & 
\hphantom{$-$}1.30 $\pm$ 1.83 & & 
10.84$\genfrac{}{}{0pt}{}{+0.02}{-0.03}$ & & 
< 0.07\hphantom{\,\footnotesize{\textsuperscript{\textdagger}}} & & 
0.18$\genfrac{}{}{0pt}{}{+0.06}{-0.05}$ & & 
0.63 $\pm$ 0.01 \\

94982 & & 
1.827 & & 
\hphantom{0}7.26 $\pm$ 0.42 & & 
$-$0.26 $\pm$ 0.67 & & 
10.96$\genfrac{}{}{0pt}{}{+0.02}{-0.01}$ & & 
< 0.09\hphantom{\,\footnotesize{\textsuperscript{\textdagger}}} & & 
0.44$\genfrac{}{}{0pt}{}{+0.03}{-0.03}$ & & 
1.28 $\pm$ 0.00 \\

106260 & & 
3.980 & & 
12.34 $\pm$ 2.12 & & 
\hphantom{$-$0}3.86 $\pm$ 13.28 & & 
10.72$\genfrac{}{}{0pt}{}{+0.05}{-0.05}$ & & 
< 0.05\hphantom{\,\footnotesize{\textsuperscript{\textdagger}}} & & 
0.23$\genfrac{}{}{0pt}{}{+0.06}{-0.05}$ & & 
1.72 $\pm$ 0.10 \\

109760 & & 
4.622 & & 
\hphantom{0}8.75 $\pm$ 0.94 & & 
\hphantom{$-$}1.77 $\pm$ 1.67 & & 
11.03$\genfrac{}{}{0pt}{}{+0.04}{-0.04}$ & & 
< 0.11\hphantom{\,\footnotesize{\textsuperscript{\textdagger}}} & & 
0.51$\genfrac{}{}{0pt}{}{+0.05}{-0.05}$ & & 
0.57 $\pm$ 0.04 \\

113667 & & 
3.971 & & 
\hphantom{0}9.21 $\pm$ 0.67 & & 
$-$0.77 $\pm$ 1.62 & & 
10.53$\genfrac{}{}{0pt}{}{+0.02}{-0.02}$ & & 
< 0.07\,\footnotesize{\textsuperscript{\textdagger}}& & 
0.16$\genfrac{}{}{0pt}{}{+0.02}{-0.04}$ & & 
0.66 $\pm$ 0.06 \\

117560 & & 
4.621 & & 
\hphantom{0}8.74 $\pm$ 0.63 & & 
$-$1.87 $\pm$ 1.32 & & 
11.08$\genfrac{}{}{0pt}{}{+0.03}{-0.03}$ & & 
< 0.46\,\footnotesize{\textsuperscript{\textdagger}} & & 
0.13$\genfrac{}{}{0pt}{}{+0.04}{-0.03}$ & & 
0.55 $\pm$ 0.03 \\

127460 & & 
2.531 & & 
\hphantom{0}7.37 $\pm$ 0.91 & & 
$-$7.63 $\pm$ 1.73 & & 
10.79$\genfrac{}{}{0pt}{}{+0.02}{-0.02}$ & & 
< 0.06\hphantom{\,\footnotesize{\textsuperscript{\textdagger}}} & & 
0.28$\genfrac{}{}{0pt}{}{+0.06}{-0.04}$ & & 
0.73 $\pm$ 0.01 \\

\hline

\end{tabular}

    \caption{Key properties of our sample. Stellar mass, star-formation rate (SFR) and time since quenching values (columns 5, 6 and 7, respectively) are derived through star-formation history modelling via \textsc{Bagpipes} (see Section \ref{sub:properties}). Negative W$_{\textrm{[O$\,$\textsc{ii}]}}$ values indicate emission. The only object classified as a quiescent galaxy, rather than a PSB, EXCELS--55410 is highlighted with an asterisk (see Section \ref{sub:sample}). Galaxy SFRs followed by \textsuperscript{\textdagger} are upper limits as given from the best-fitting \textsc{Bagpipes} star-formation histories. All other SFR limits are quoted as the minimum SFR needed for a specific star-formation rate of log$_{10}$(sSFR) = $-$12, roughly the smallest sSFR that can be detected \citep[e.g.,][]{fang_demographics_2018, belli_mosfire_2019, carnall_vandels_2019}. The upper limit obtained from the \textsc{Bagpipes} fitting is substantially lower than the quoted value for these galaxies, due to the use of our double-power-law SFH model.}
    \label{tab:props}
\end{table*}

\endgroup

Some studies include an additional PSB selection criterion, using the [O$\,$\textsc{ii}] ($\lambdaup\lambdaup$\,3727,\,3729\,\AA) equivalent width, W$_{\textrm{[O$\,$\textsc{ii}]}}$, to remove any possible star-forming contaminants from their sample \citep[e.g.,][]{tran_nature_2003, maltby_identification_2016}. We do not impose this W$_{\textrm{[O$\,$\textsc{ii}]}}$ cut on objects selected with W$_{\textrm{H$\delta$}}$ > 5\,\AA, to avoid biasing against PSBs potentially hosting AGN. We do, however, use the [O$\,$\textsc{ii}] equivalent width to search for and select older passive galaxies from the four objects in our parent sample with W$_{\textrm{H$\delta$}}$ < 5\,\AA, requiring W$_{\textrm{[O$\,$\textsc{ii}]}}$ > $-5$\,\AA\ \citep[see, e.g.,][]{maltby_high-velocity_2019}. To calculate W$_{\textrm{[O$\,$\textsc{ii}]}}$, we use Equation \ref{eq:ew}, with $\lambda_{1} = 3713$\,\AA, $\lambda_{2} = 3741$\,\AA, and estimate $F_{c}(\lambda)$ using the wavelength ranges 3653\,--\,3713 and 3741\,--\,3801\,\AA. Three of the four parent sample galaxies not selected as PSBs do not have coverage of the [O$\,$\textsc{ii}] line, and we therefore remove them, as we cannot effectively rule out the possibility of ongoing star formation from the available spectroscopy\footnote{Some studies use a cut in the H$\mathrm{\alpha}$ equivalent width to select `pure' PSBs \citep[or E+A galaxies; see e.g.,][]{wilkinson_evolutionary_2017}, however this additionally selects against PSBs hosting AGN (as well as residual star formation and shocks), which we attempt to avoid in this work.}. The remaining galaxy from our parent sample meets the passive galaxy criteria (EXCELS--55410), and we include this object in the overall sample of PSBs and passive galaxies used in this work. The equivalent width measurements of our final sample of 13 PSB and passive galaxies can be found in Table \ref{tab:props}. 

We compare our sample to the commonly used rest-frame UVJ selection method  \citep[][]{williams_detection_2009, muzzin_evolution_2013}. We note that this is for comparison purposes only, and is not part of our sample selection process. The selection of quiescent galaxies using this method has been shown to hold out to $z$ = 3.75 in \cite{carnall_inferring_2018}, which is the median $z_{\mathrm{spec}}$ of our sample. We calculate the rest-frame UVJ colours of our galaxies using \textsc{Bagpipes} (see Section \ref{sub:properties} for more detail), and these are presented in Fig. \ref{fig:uvj}. Our PSBs selected with W$_{\textrm{H$\delta$}} > 5$\,\AA\ are shown in black, and the older quiescent galaxy is shown as a magenta triangle. The solid diagonal line shows the \cite{muzzin_evolution_2013} quiescent selection criteron ($U - V >  0.88(V - J) + 0.59$) for galaxies with redshifts in the range 1 $< z <$ 4, and the shaded blue region shows a post-starburst selection box for galaxies with a median stellar age of 0.3 $< t_{50} <$ 1\,Gyr, as given in \cite{belli_mosfire_2019}\footnote{$t_{50} = 7.03 + 0.83(V - J) + 0.74(U - V)$}. Three of our galaxies lie just outside the quiescent region of the diagram, two of which are X--ray detected (EXCELS--34495 and EXCELS--106260, see Section \ref{sub:drivers} for further discussion). \cite{stevenson_primer_2025} compared quiescent galaxies selected based on sSFR and UVJ cuts, and found very good agreement, with only slight discrepancies in the `green valley' area, just below the quiescent galaxy selection box. Our sample is consistent with their results.

In summary, we obtain a final sample of 13 galaxies, selected with the following criteria:
\begin{itemize}
    \item EXCELS galaxies initially selected for observation with a high likelihood of being passive based on photometric fitting (see Section \ref{sub:EXCELS}),
    \item Full coverage of the H$\delta\, (\lambdaup$\,4102\,\AA) and Na\,\textsc{d} features,
    \item W$_{\textrm{H$\delta$}}$ > 5\,\AA\ (PSB) or W$_{\textrm{H$\delta$}}$ < 5\,\AA\ and W$_{\textrm{[O$\,$\textsc{ii}]}}$ > $-5$\,\AA\ (passive).
\end{itemize}

\subsection{Literature Comparison Sample}\label{sec:compsamp}

To place our results in a wider context, we draw a sample of comparable objects from recent studies, focusing on galaxies at cosmic noon and beyond ($z \gtrsim 1.5$). We draw the comparison sample from literature in which outflow velocities are determined via analysis of Na$\,$\textsc{d}, to minimise any systematics that could arise from including studies employing different absorption lines. A comparison of the different selection techniques employed by these different works can be found in Appendix \ref{appendix}. We compare outflow incidence rates reported from the high-redshift sample, and an additional low-redshift sample, in Section \ref{section:results}. The high-redshift comparison sample is fully utilised in Section \ref{section:propertycomparison} onwards. 

\subsubsection{High-redshift}\label{subsub:hz}

The Blue Jay survey \citep[][]{belli_blue_2025} detected outflows in the sodium absorption profiles of six quiescent galaxies at cosmic noon \citep[1.7 $< z <$ 3.5, \citetalias{davies_jwst_2024};][]{park_widespread_2024}. We include these in Fig. \ref{fig:tsq} onwards. For completeness, we also include those galaxies selected as quiescent from Blue Jay without an outflow detection. In total, the Blue Jay quiescent sample with coverage of Na$\,$\textsc{d} (hereinafter referred to as Blue Jay QG) comprises 13 galaxies, with stellar masses $10.3 \leq \mathrm{log}_{10}(M_{*}/\mathrm{M}_{\odot}) \leq 11.7$, redshifts $1.7 \leq z_{\mathrm{spec}} \leq 2.6$, and star formation rates $-2.5 \leq \mathrm{log}_{10}(\mathrm{SFR / \mathrm{M_{\odot}\,yr}^{-1}}) \leq 0.5$. A comparison of our selection method (as outlined in Section \ref{sub:sample}) and the Blue Jay QG selection method is presented in Appendix \ref{appendix}.

We also include two more sources, taken from studies focusing on one or two rapidly quenched (quenching timescales $\lesssim 200\,$Myr) galaxies in detail. The first is JADES-GS-206183, as presented in \cite{sun_extreme_2026}, at $z = 1.317$. The second object is NS\_274, at $z = 4.106$: outflow properties in this galaxy were originally studied using the Mg$\,$\textsc{ii} ($\lambdaup\lambdaup$\,2796,\,2803\,\AA) absorption profile in \cite{wu_ejective_2025}, and subsequent work by \cite{valentino_gas_2025} extended the analysis to include Na$\,$\textsc{d}.\footnote{\cite{valentino_gas_2025} present an additional recently quenched object in their work, however their analysis does not extend to Na$\,$\textsc{d} for this object, thus we do not include it in our sample.} 

\subsubsection{Low-redshift}\label{sub:lowz}

We also compare our results to the work of \citetalias{sun_evolution_2024}, who studied a sample of $\sim 500$ PSBs from SDSS. We stress that direct comparison between \citetalias{sun_evolution_2024} and our sample is not simple, due to different selection criteria, spectral resolutions, and methodologies employed to model the Na$\,$\textsc{d} profile. While we use a model based on physical parameters, \citetalias{sun_evolution_2024} use a more straightforward parameterisation. We refer the reader to their study for more details of the fitting procedure. Outflow velocities in \citetalias{sun_evolution_2024} are given as the average velocity offset from the galaxy redshift for the entire PSB population, and thus we do not compare our velocity values with theirs, as our measurements are for individual galaxies (see Section \ref{sub:fitting} for full details of how we fit the Na$\,$\textsc{d} profile). The main factor cautioning against direct comparison between \citetalias{sun_evolution_2024} and our sample is the PSB selection criteria. \citetalias{sun_evolution_2024} select PSBs based on the Lick H$\delta_\mathrm{A}$ absorption line index and H$\alpha$ equivalent width, requiring H$\delta_{\mathrm{A}}$ > 4\,\AA\ and W$_{\textrm{H$\alpha$}}$ < 3\,\AA. These criteria select against strong AGN, ongoing residual star formation, and shocks, which we do not do in this work.

\section{Methods and Analysis}\label{section:analysis}

\subsection{Star-formation History Modelling}\label{sub:properties}

We model the star-formation histories (SFHs) of galaxies in our sample using \textsc{Bagpipes} to simultaneously fit the NIRSpec spectra and the PRIMER photometric data, using the catalogue described in \cite{begley_evolution_2025}. It should be noted that this is a separate fitting process to the \textsc{Bagpipes} fits to photometry performed for spectroscopic calibration in Section \ref{sub:EXCELS}. The SFH fitting in this work follows the prescription of \cite{carnall_jwst_2024}, which we briefly summarise here.

The spectroscopic data for each galaxy in our sample is restricted to 3450\,--\,7350\,\AA\ (rest-frame), as this is the range spanned by the MILES stellar spectral library \citep[][]{falcon-barroso_updated_2011} within the 2016 updated \cite{bruzual_stellar_2003} models \citep[see][]{chevallard_modelling_2016}. We fit SFHs using a double power-law, with the dust attenuation model of \cite{salim_dust_2018}. A 38$^{\mathrm{th}}$--order polynomial is used to correct for any imperfections in the spectrophotometric calibration \citep[see, e.g.,][]{carnall_jwst_2024}, with a `white-scaled' noise component to account for underestimated uncertainties in the spectra. The noise component is allowed to vary from $\alpha$ = 1\,--\,10, with a logarithmic prior. Velocity dispersion (beyond that inherent to the SSP models) is modelled by convolution of the model spectrum with a Gaussian kernel with width $\sigma$, which we allow to vary between 50\,--\,500\,km\,s$^{-1}$, with a logarithmic prior. Each galaxy spectrum is assigned a mask, excluding emission lines [O$\,$\textsc{ii}], [O$\,$\textsc{iii}], [N$\,$\textsc{ii}] and [S$\,$\textsc{ii}], and the Na$\,$\textsc{d} absorption feature. We fix the ionization parameter to log$_{10}(U) = -3$, as masking the aforementioned spectral emission lines means the nebular component will contribute a negligible amount to the model spectrum. One object, EXCELS--34495, displays a clear broad H$\alpha$ line, as is the case for the similar object, GS-9209, described in \cite{carnall_massive_2023}. We therefore include the \textsc{Bagpipes} AGN component described in that work for object EXCELS--34495 only. The full list of model parameters and priors is given in Table \ref{tab:bpparam}.

\begin{table*}
    \centering
        \begin{tabular}{llccc}
    \hline
        Component & Parameter & Range & Prior & Hyper-parameters \\
    \hline
    \hline

            & log$_{10}({M}_{*}/\mathrm{M}_{\odot}$) & (1, 13) & Logarithmic & \\
            & $Z_{*}$ ($Z_{\odot}$) & (0.1, 5) & Logarithmic & \\
        SFH & Turn-over time, $\tau$ (Gyr) & (0.1, $t_{\mathrm{obs}}$)\textsuperscript{\textdagger} & Uniform & \\
            & Falling slope, $\alpha$ & (0.1, 1000) & Logarithmic & \\
            & Rising slope, $\beta$ & (0.1, 1000) & Logarithmic & \\
    
    \hline

             & $A_{V}$ (mag) & (0, 4) & Uniform & \\
        Dust & Deviation from Calzetti slope, $\delta$ & (-0.3, 0.3) & Gaussian & $\mu$ = 0, $\sigma$ = 0.1 \\
             & Strength of 2175\AA \hspace{1pt} bump, $B$ & (0, 5) & Uniform & \\

    \hline

             & $z$ & $z_{\mathrm{spec}} \, \pm$ 0.05 & Gaussian & $\mu$ = $z_{\mathrm{spec}}$, $\sigma$ = 0.1 \\
        Other & Velocity dispersion, $\sigma$ (km\,s$^{-1}$) & (50, 500) & Logarithmic & \\
             & White noise scaling, $\alpha$ & (0.1, 10) & Logarithmic & \\
             \hline

& Power law slope, $\alpha_{\lambda < 5000\mathrm{\AA}}$ & ($-2$, $2$) & Gaussian & $\mu=-1.5$, $\sigma=0.5$ \\
& Power law slope, $\alpha_{\lambda > 5000\mathrm{\AA}}$ & ($-2$, $2$) & Gaussian & $\mu=0.5$, $\sigma=0.5$ \\
AGN (EXCELS--34495 only) & $f_{\mathrm{H}\alpha\mathrm{,\,broad}}$ (erg\,s$^{-1}$\,cm$^{-2}$) & (0, $2.5\times10^{-17}$) & Uniform \\
& $\sigma_{\mathrm{H}\alpha\mathrm{,\,broad}}$ (km\,s$^{-1}$) & (1000, 5000) & Logarithmic & \\
& $f_{5100}$ (erg\,s$^{-1}$\,cm$^{-2}$\,\AA$^{-1}$) & (0, $10^{-19}$) & Uniform & \\
    \hline
    \multicolumn{5}{l}{\multirow{2}{40em}{\footnotesize{\textsuperscript{\textdagger} $t_{\mathrm{obs}}$ is the age of the Universe at the spectroscopic redshift of observation.}}} \\ \\
    
    \end{tabular}
    \caption{Parameters and prior distributions used for the \textsc{Bagpipes} star-formation history modelling (Section \ref{sub:properties}). Logarithmic priors are given in base 10.}
    \label{tab:bpparam}
\end{table*}

We then derive some quantities of interest from the fitted SFH models. \textsc{Bagpipes} defines a galaxy as quenched using the specific star-formation rate (sSFR) criterion of sSFR < 0.2 / $t_{\mathrm{H}}(z)$ (see Section \ref{sub:EXCELS}). From the fitted SFH models, \textsc{Bagpipes} provides the age of the Universe at the time of quenching, $t_{\mathrm{quench}}$. We define the time \textit{since} quenching, $t_{\mathrm{sq}}$ as:
\begin{equation}\label{eq:tsq}
    t_{\mathrm{sq}} = t_{\mathrm{obs}} - t_{\mathrm{quench}}, 
\end{equation}
where $t_{\mathrm{obs}}$ is the age of the Universe at $z_{\mathrm{spec}}$. Derived properties from the SFH modelling are presented in Table \ref{tab:props}. Our sample spans stellar masses 10.4 $\leq \mathrm{log}_{10}(M_{*}/\mathrm{M}_{\odot}) \leq$ 11.1, redshifts 1.8 $\leq z_{\mathrm{spec}} \leq$ 4.6, and star-formation rates $\mathrm{log}_{10}(\mathrm{SFR / \mathrm{M_{\odot}\,yr}^{-1}}) < -0.3$.

\subsection{Galaxy Effective Radii}\label{sub:structure}

In this work, we make use of galaxy effective radius measurements, $r_{\mathrm{eff}}$, determined in the PRIMER-UDS F200W band. Full details of the process used to measure $r_{\mathrm{eff}}$ are presented in Maltby et al. (in prep), and we briefly outline them here.

Maltby et al. run the {\sc{galapagos}}{\scriptsize{-2}} pipeline \citep{hausler_galapagos-2galfitmgama_2022} on each PRIMER-UDS NIRCam waveband. Sources are identified via {\sc{SExtractor}} \citep{bertin_sextractor_1996}, and a postage stamp is extracted for each. Residual sky background is estimated on a galaxy by galaxy basis, and an additional local background subtraction is applied. An empirical PSF is determined directly from the imaging, via stacking of isolated stars within the PRIMER-UDS field.

To determine structural parameters, {\sc{galapagos}}{\scriptsize{-2}} employs {\sc{galfitm}} \citep[][]{hausler_megamorph_2013} to fit PSF-convolved single-component S\'ersic profiles to each source, of the form:

\begin{equation}
I(r) = I_{\rm eff} \exp \left\{ -b_n \left[ \left( \frac{r}{r_{\rm eff}} \right)^{1/n} - 1 \right] \right\}.
\end{equation}
Here, $I_{\rm eff}$ is the intensity at the effective radius, $n$ is the S\'ersic index, and $b_n$ is a normalization constant dependent on $n$. 

\subsection{Fitting NaD Absorption Profiles}\label{sub:fitting}

In this sub-section, we outline the method used to fit the Na$\,$\textsc{d} absorption profiles in our galaxies. We first model and remove any stellar contribution to the absorption feature (detailed in Section \ref{subsub:ppxf}). The residual flux is then fitted with a model based on physical properties of the gas (detailed in Section \ref{subsub:model}), and the presence of gas flows is classified (Section \ref{subsub:classification}).

\subsubsection{Modelling the Stellar Component}\label{subsub:ppxf}

The Na$\,$\textsc{d} absorption profile may have contributions from both systemic stellar and ISM components. To model and remove any stellar absorption, we fit each galaxy spectrum using the E-MILES stellar population synthesis (SPS) models \citep[][]{vazdekis_uv-extended_2016} via the Penalized PiXel-Fitting \citep[\textsc{ppxf};][]{cappellari_full_2023} software. The fits are performed over the wavelength range 3540\,--\,7350\,\AA, and we mask several spectral lines that may affect the stellar modelling, as in Section \ref{sub:properties}. We note that \textsc{ppxf} allows simultaneous modelling of ionized gas emission, however when tested this had no significant effect on our results. We allow a fourth-degree multiplicative polynomial to correct for any mismatches between the templates and the galaxy spectrum. The code outputs a best-fitting redshift, and we confirm that these agree very well with the initial input redshifts taken from \cite{carnall_jwst_2024}. 

We note that the chosen SPS library used within \textsc{ppxf} may predict different stellar Na\,\textsc{d} absorption profile strengths. We therefore test this effect by re-running our \textsc{ppxf} fits (and subsequent analysis, described in the following sections) using the FSPS \citep[][]{conroy_propagation_2009, conroy_propagation_2010}, GALAXEV \citep[][]{bruzual_stellar_2003} and XSL \citep[][]{verro_modelling_2022} libraries. We find the choice of SPS library has no significant effect on our results, with all inflow, outflow, and non-detection rates remaining the same.

\begin{figure*}
    \centering
    \includegraphics[width=\textwidth]{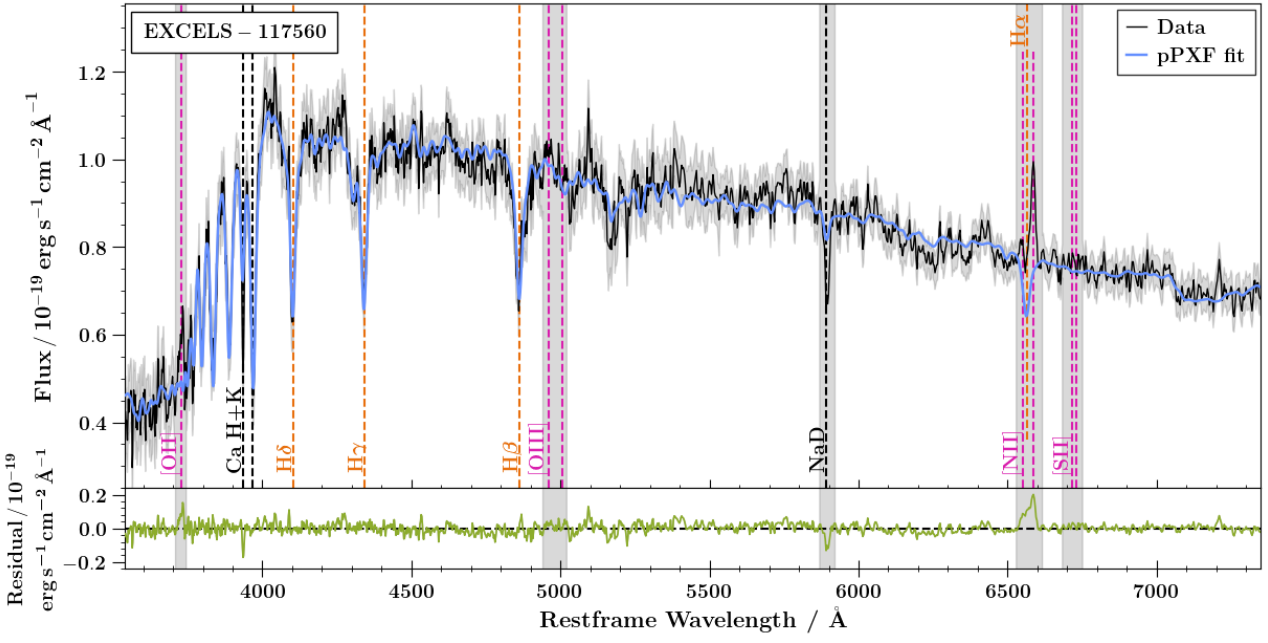}
    \caption{Example \textsc{ppxf} fit (blue, see Section \ref{sub:fitting}) to a galaxy spectrum in our sample (black). The rest-frame wavelengths of emission and absorption lines of interest are marked with vertical dotted lines. The shaded grey area shows the error on the flux, and features that are masked during fitting are denoted by shaded grey sections spanning the full y-axis. Within our sample, EXCELS--117560 is a relatively high-redshift, relatively massive galaxy ($z_{\mathrm{spec}}=4.62$, $\mathrm{log}_{10}(M_{*}/\mathrm{M}_{\odot}) \simeq 11.1$). Residuals between the \textsc{ppxf} fit and data are shown below the main panel in green.}
    \label{fig:ppxf}
\end{figure*}

We calculate the equivalent width of the Na$\,$\textsc{d} profile prior to removal of the stellar component, W$_{\textrm{Na\,\textsc{d}}}$, using Equation \ref{eq:ew} with $\lambda_{1}$ and $\lambda_2$ values of 5870 and 5920 \AA, respectively. We use the best-fitting stellar continuum model to estimate $F_c$, with continuum intervals 5850 -- 5870 and 5920 -- 5940\,\AA. Prior to more detailed fitting of the Na$\,$\textsc{d} profile, each galaxy spectrum is normalised by dividing through by the best-fitting \textsc{ppxf} model to remove any stellar continuum contribution. We refer to the stellar-continuum-normalised spectrum as the `residual flux' throughout the remainder of the paper. An example \textsc{ppxf} fit to one of the galaxies in our sample is shown in Fig. \ref{fig:ppxf}.

\subsubsection{Modelling Residual Flux}\label{subsub:model}

The He$\,$\textsc{i}$\,\lambda$\,5875\,\AA\ emission line lies in close proximity to the Na$\,$\textsc{d} absorption lines, and recent studies (e.g., \citetalias{davies_jwst_2024}) have thus simultaneously fitted absorption and emission lines to account for possible infilling of the profile. On inspection, we find no clear evidence for He$\,$\textsc{i} emission in our spectra, and we elect to fit the Na$\,$\textsc{d} absorption profile only. We note that following the method of \citetalias{davies_jwst_2024} has no significant effect on our results.

We fit the residual flux, $F_{\mathrm{Na\,\textsc{d}}}(\lambda_i)$, of each spectrum between 5820\,--\,5970\,\AA\, using the standard partial covering model of \cite{rupke_outflows_2005}:
\begin{equation}
    F_{\mathrm{Na\,\textsc{d}}}(\lambda_i) = 1 - C_f + C_f \;e^{(-\tau_{b}(\lambda_i) - \tau_r(\lambda_i))},
\end{equation}
where $C_f$ is the gas covering fraction, and $\tau_b$ and $\tau_r$ are the optical depth profiles of the 5891 and 5897\,\AA\ Na$\,$\textsc{d} lines, respectively, given by:
\begin{align}
        \tau_b(\lambda_i) &= 2\tau_0\,e^{- (\lambda_i - \lambda_c)^2 \,/\,2\sigma^2}, \\
        \tau_r(\lambda_i) &= \tau_0\,e^{- (\lambda_i - \lambda_c\,-\,6)^2 \,/\,2\sigma^2}.
\end{align}
Here, $\lambda_c = (5891\,$\AA$\,-\,\Delta\lambda)$, is the offset of the Na$\,$\textsc{d} profile from its rest-frame wavelength, where $\Delta\lambda = [-10,10]\,$\AA, and $\tau_0$ is the optical depth at the centre of the 5897\,\AA\ line, fixed to be half that of the 5891\,\AA\ line. This line ratio is suitable for optically thin gas \citep{draine_physics_2011}, and we find that allowing the line ratio to vary does not significantly affect our results. We convolve the line width of each absorption line, $\sigma_{\mathrm{Na\,\textsc{d}}}$, with the NIRSpec instrumental dispersion, $\sigma_{\mathrm{inst}}$  prior to fitting, such that $\sigma = \sqrt{\sigma_{\mathrm{Na\,\textsc{d}}}^2 + \sigma_{\mathrm{inst}}^2}$. We determine $\sigma_{\mathrm{inst}}$ at the wavelength of observation of Na$\,$\textsc{d} for each galaxy, using the dispersion curves from the JWST User Documentation (JDox)\footnote{\href{https://jwst-docs.stsci.edu/jwst-near-infrared-spectrograph/nirspec-instrumentation/nirspec-dispersers-and-filters}{https://jwst-docs.stsci.edu/jwst-near-infrared-spectrograph/nirspec-instrumentation/nirspec-dispersers-and-filters}}. Recent work has found that the NIRSpec resolution for compact sources is higher than stated \citep[][]{de_graaff_ionised_2024}, and we thus scale the initial instrumental dispersion values up by a factor of 1.3 \citep[e.g.,][]{killi_deciphering_2024, greene_uncover_2024}. At $z_{\mathrm{spec}}=3.8$, the median redshift of our sample, the JDox instrumental resolution at the (rest-frame) wavelength of Na\,\textsc{d} is $\approx3.7$\AA, corresponding to $\sigma_{\mathrm{inst}}\approx2.9$\AA\ (\textit{R} $\sim$ 2000), after our scaling.

In total, we fit four free parameters: $C_f$, $\tau_0$, $\sigma$ and $\lambda_c$. We perform our fits using Bayesian analysis via \textsc{Nautilus} \citep[][]{lange_nautilus_2023}, to provide accurate estimations of our parameter uncertainties and degeneracies. This approach also helps to account for the degeneracy between $C_f$ and $\tau_0$, which is present when the Na$\,$\textsc{d} lines are blended, such as in our data. The best-fitting model uses the median values of the posterior probability distributions of each parameter, and uncertainties are calculated using the 16th\,--\,84th posterior percentile range.

\subsubsection{Classifying Gas Flows}\label{subsub:classification}

We determine whether a galaxy has excess absorption of Na$\,$\textsc{d} after removing the stellar continuum by calculating the equivalent width of the best-fitting model, W$_{\textrm{ISM}}$. Uncertainties are calculated using the full posterior probability distributions of our model parameters. Galaxies are classified as having excess absorption if W$_{\textrm{ISM}}$ is detected at $> 5\sigma$, otherwise we term them as having `no residual'.

For galaxies in which we determine excess absorption is present, we determine the velocity offset of the best-fitting absorption profile
\begin{equation}\label{eq:dv}
    \Delta v = \frac{\Delta\lambda}{5891\,\text{\AA}}\,c\,,
\end{equation}
where $\Delta\lambda = (5891$\,\AA$\,-\, \lambda_c)$ (see Section \ref{subsub:model}), and $c$ is the speed of light. Where the velocity offset is consistent with the galaxy systemic velocity within the uncertainties ($\Delta v = 0$\,km\,s$^{-1}$), the absorption likely originates in neutral gas within the ISM, tracing the bulk motion of the galaxy. If a galaxy has a strongly blueshifted or redshifted absorption profile (i.e. $\Delta v \neq 0$\,km\,s$^{-1}$, detected at $>1\sigma$), this is a sign of an outflow or inflow. In this work, we denote outflowing gas with a positive $\Delta v$ value, and inflowing gas has a negative $\Delta v$. For these objects, we calculate the gas flow velocity, $v_{\mathrm{flow}}$, using
\begin{equation}\label{eq:vflow}
    v_{\mathrm{flow}} = |\Delta v| + 2\sigma_{\mathrm{Na\,\textsc{d}}} \,.
\end{equation}
This definition of $v_{\mathrm{flow}}$ assumes that the gas flow is seen at all inclinations, and therefore much of the gas has a velocity component perpendicular to our line of sight. This definition has been used throughout recent high-redshift Na$\,$\textsc{d} studies \citep[e.g., \citetalias{davies_jwst_2024};][]{belli_star_2024, sun_extreme_2026}. In this case, the intrinsic gas flow velocity ($v_{\mathrm{flow}}$) corresponds to gas flow along the line of sight. Uncertainties on $v_{\mathrm{flow}}$ are calculated using the full posterior distributions for $\Delta v$ and $\sigma_{\mathrm{Na\,\textsc{d}}}$ outputted from \textsc{Nautilus}. The calculated W$_{\textrm{Na\,\textsc{d}}}$, W$_{\textrm{ISM}}$, $\Delta v$ and $v_{\mathrm{flow}}$ values are reported in Table \ref{tab:results}. We note that the choice of SPS model used (see Section \ref{subsub:ppxf}) has no significant effect on our W$_{\textrm{Na\,\textsc{d}}}$, W$_{\textrm{ISM}}$, and $v_{\mathrm{flow}}$ results. For outflow and inflow detected galaxies, W$_{\textrm{Na\,\textsc{d}}}$, W$_{\textrm{ISM}}$ and $v_{\mathrm{flow}}$ values vary no more than 15 per cent, and are all consistent within the errors. For the remaining galaxies, W$_{\textrm{Na\,\textsc{d}}}$ and W$_{\textrm{ISM}}$ vary by up to 40 per cent, however these all remain non-detections (W$_{\textrm{ISM}}$ not detected at $> 5\sigma$), and all values are again consistent within the errors.

\section{Results}\label{section:results}

\begin{figure*}
    \centering
    \includegraphics[width=\textwidth]{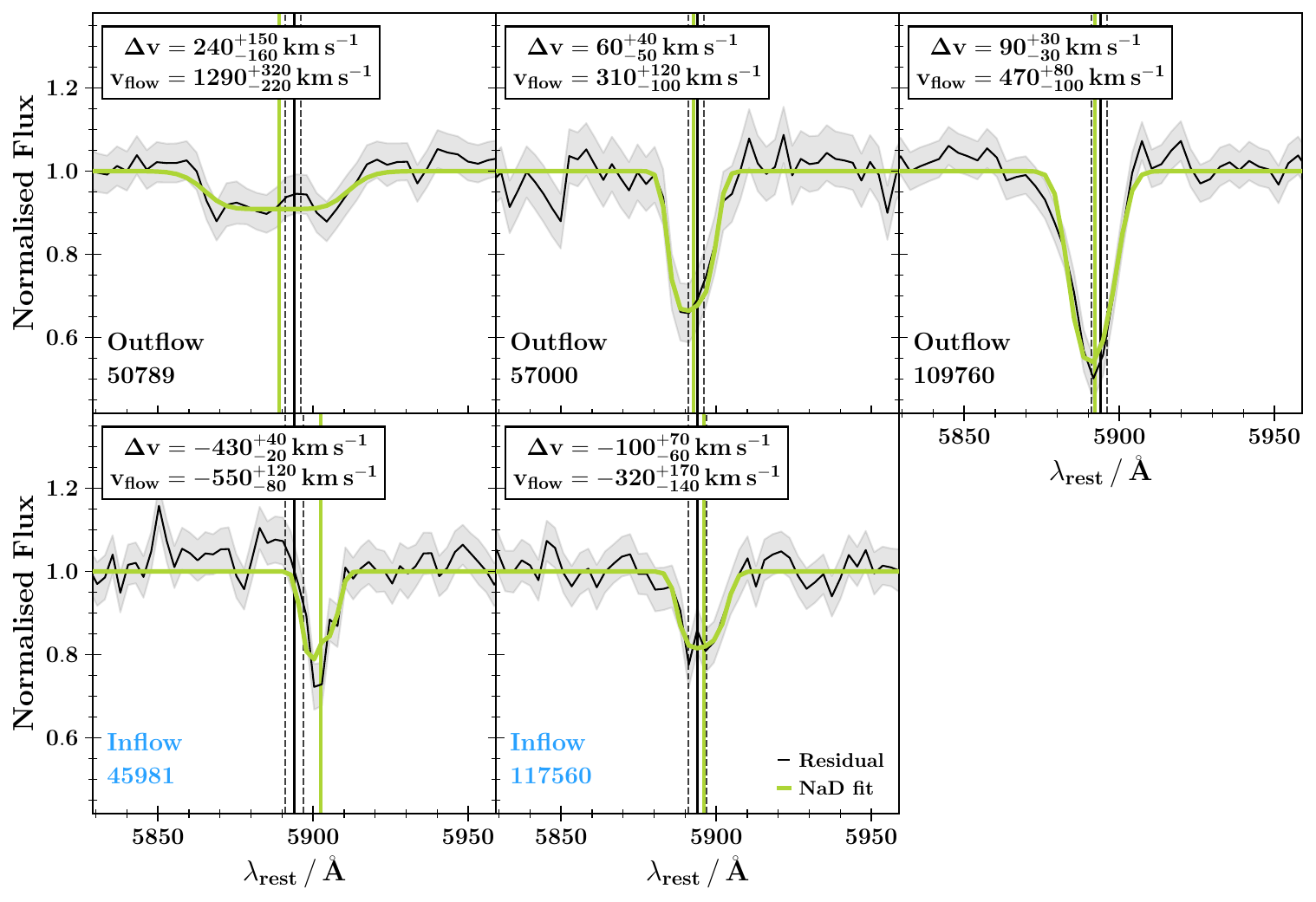}
    \caption{Best-fitting Na$\,$\textsc{d} models (green) for the five galaxies in our sample that have signs of outflowing (black labels, top panels) or inflowing (blue labels, bottom panels) gas. The fitting is performed on the residual flux (black) after each spectrum has been normalised by the best-fitting stellar continuum model from \textsc{ppxf} (see Section \ref{subsub:ppxf}). Solid black vertical lines for each panel denote the central wavelength of the Na$\,$\textsc{d} feature (the individual line wavelengths $\lambdaup\lambdaup$\,5891,\,5897\,\AA\ are marked with dashed black vertical lines). Solid green vertical lines denote the best-fitting central wavelength of our fitted models. All $\Delta v$ values and associated uncertainties are quoted in km\,s$^{-1}$ (see Equation \ref{eq:dv}). The residual Na$\,$\textsc{d} profiles of the remaining objects in our sample are shown in Fig. \ref{fig:nadnoflow}. The best-fitting models indicate that $23^{+13}_{-10}$ per cent of our sample show signs of outflows, with a further $15^{+12}_{-7}$ per cent showing signs of inflows. }
    \label{fig:nadflow}
\end{figure*}

\begin{figure*}
    \centering
    \includegraphics[width=\textwidth]{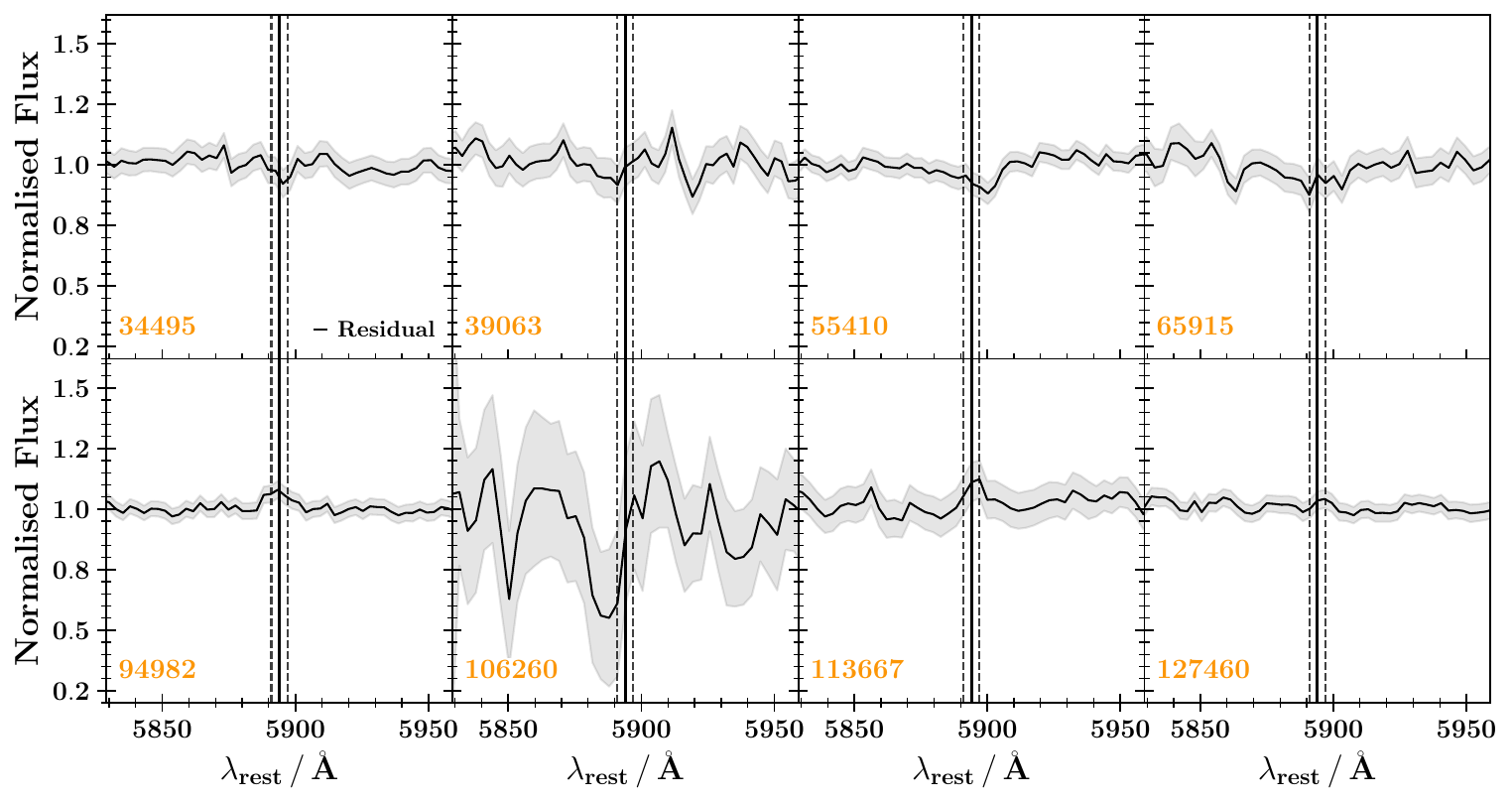}
    \caption{Residual Na$\,$\textsc{d} profiles for the remaining eight galaxies in our sample. These galaxies were deemed to have insubstantial residual flux to fit our model robustly, once the stellar continuum was removed. Vertical lines are as described in Fig. \ref{fig:nadflow}. Over half our sample have residual Na$\,$\textsc{d} profiles that indicate no sign of excess absorption after accounting for the stellar component (see Section \ref{subsub:excess}).}
    \label{fig:nadnoflow}
\end{figure*}

The EXCELS spectroscopic data, reduced as described in Section \ref{sub:EXCELS} and continuum normalised as described in Section \ref{subsub:model}, are shown in Figs \ref{fig:nadflow} and \ref{fig:nadnoflow}, along with the best-fitting residual flux models, fitted as described in Sections \ref{subsub:model} and \ref{subsub:classification}. Fig. \ref{fig:nadflow} shows galaxies with evidence for gas outflows or inflows, and Fig. \ref{fig:nadnoflow} shows galaxies with no evidence for gas flows. We find that a range of Na$\,$\textsc{d} profiles are present in our sample, and we discuss these in the following sections.

\subsection{Evidence for Gas Flows}\label{sec:gf}

We find that three galaxies ($23^{+13}_{-10}$ per cent) have signs of blueshifted absorption indicating outflows (top row in Fig. \ref{fig:nadflow}). Uncertainties in the detection rate are the 1$\sigma$ limits derived from 10 million simulated binomial populations of 13 `trials' with a rate parameter of 3 `successes' (outflows). The same approach is used to calculate confidence intervals throughout this section. We note that two galaxies in Fig. \ref{fig:nadflow} (EXCELS--57000 and 117560) have velocities very close to the Na\,\textsc{d} systemic velocity, with velocity shifts only detected at a $\simeq1.5\sigma$ confidence level. While we still classify these as outflows/inflows for consistency with recent literature, it is possible that these galaxies are showing ISM gas at systemic velocity, rather than galactic winds. The resolution of our data prevents us from being able to confidently separate the Na\,\textsc{d} absorption profiles into ISM and outflow/inflow components, and thus highlights the need for higher-resolution spectroscopy of these sources. With this in mind, we find that our detection rate is consistent with that of the Blue Jay QG sample (6/13, $46^{+14}_{-13}$ per cent; see Section \ref{subsub:hz}) to a level of $1.3\sigma$. This is calculated via P($\mathcal{N}(46, 13.5^2) - \mathcal{N}(23, 11.5^2) < 0)$, given by $(46-23)/\sqrt{(13.5)^2 + (11.5)^2}=1.3\sigma$.

At low redshift, \citetalias{sun_evolution_2024} found $\sim20$ per cent of PSBs in their SDSS sample had outflowing gas, along with \cite{baron_multiphase_2022}, who found $\sim20$ per cent of PSBs hosting AGN showed signatures of neutral outflows. Similarly, \cite{coil_outflowing_2011} report an outflow detection rate in PSBs of $\sim30$ per cent, when using Mg$\,$\textsc{ii} and Fe$\,$\textsc{ii} absorption profiles as tracers of galactic winds. As noted in Section \ref{sub:lowz}, PSB selection techniques vary between studies, and may exclude AGN hosts, making a direct comparison less straightforward. Despite this, it is encouraging to see that our results are consistent with those found at low redshift.

We find modest wind velocities in two out of three of our galaxies with detected outflows ($\Delta v \leq 100\,$ km\,s$^{-1}$, $v_{\mathrm{flow}} \leq 500\,$ km\,s$^{-1}$), while EXCELS--50789 has a much higher velocity outflow ($\Delta v \sim 240\,$ km\,s$^{-1}$, v$_{\mathrm{flow}} >$ 1250\,km\,s$^{-1}$), comparable to winds observed in PSBs at cosmic noon \citep[e.g.,][]{tremonti_discovery_2007, taylor_high-velocity_2024}. We discuss this object in further detail in Section \ref{subsub:50789}. The range of outflow velocities in this work is consistent with the Blue Jay QG sample ($v_{\mathrm{flow,\,Blue\,Jay\,QG}} \sim 300-1100$\,km\,s$^{-1}$).

Two galaxies ($15^{+12}_{-7}$ per cent) have redshifted absorption, indicating inflowing gas (panels with blue labels in Fig. \ref{fig:nadflow}). Within the full Blue Jay sample, three galaxies ($\sim3$ per cent) show signs of inflowing gas, with none falling in the Blue Jay QG sample. \citetalias{sun_evolution_2024} report an inflow detection rate of $\sim10$ per cent in their PSB sample, consistent with our results. The authors additionally find that objects with inflows have smaller residual W$_{\textrm{Na\,\textsc{D}}}$ values than those with outflows, and we confirm the same in this work (see Table \ref{tab:results}). The inflow velocities are moderate, with , $v_{\mathrm{flow}} \sim 550$\,km\,s$^{-1}$ ($\Delta v \sim -430$\,km\,s$^{-1}$) and $v_{\mathrm{flow}} \sim 300$\,km\,s$^{-1}$ ($\Delta v \sim -100$\,km\,s$^{-1}$) for EXCELS--45981 and EXCELS--117560 respectively.

The remaining eight galaxies in our sample ($62^{+12}_{-14}$ per cent) are determined to have no excess Na\,\textsc{d} absorption beyond the stellar component (Fig. \ref{fig:nadnoflow}). One object in the Blue Jay QG sample ($8^{+11}_{-5}$ per cent) was deemed to not have a Na$\,$\textsc{d} detection beyond the stellar contribution (this is discussed in Section \ref{subsub:excess}). At lower redshift non-detection rates were reported as 52 per cent and 43 per cent for \cite{baron_multiphase_2022} and \citetalias{sun_evolution_2024}, respectively, consistent with our sample.

\subsubsection{Lack of Excess Na\,\textsc{d} Absorption at Systemic Velocity}\label{subsub:excess}

Formally, we find no galaxies with excess absorption at systemic velocity after accounting for the stellar component (but see the previous section for a discussion of two objects with very small velocity offsets). Six of the galaxies in the Blue Jay QG sample have excess absorption at systemic velocity ($46^{+14}_{-13}$ per cent), while \citetalias{sun_evolution_2024} found that $\sim7$ per cent of their \textit{PSBs} had excess absorption at systemic velocity. When excess sodium absorption is detected in older, passive galaxies, it is commonly found at the galaxy systemic velocity \citep[e.g.,][]{concas_two-faces_2019, carnall_stellar_2022}. The difference in the Blue Jay QG and \citetalias{sun_evolution_2024} results, along with our lack of detections of excess absorption at systemic velocity, may be due to variations in sample selection, or simply a result of the small Blue Jay QG sample size.  

The Blue Jay QG sample is lower redshift ($1.7 \leq z_{\mathrm{spec}} \leq 2.6$) than our sample, and selected on sSFR alone, with no spectroscopic cuts to distinguish younger, more recently quenched galaxies (such as PSBs) from older passive galaxies. It is possible that the Blue Jay QG sample is therefore composed of older galaxies, observed longer after the mechanism that drove any outflowing winds in these objects. 

The Blue Jay QG sample also contains galaxies at higher mass than the EXCELS sample analysed in this work (stellar mass ranges are $\mathrm{log}_{10}(M_{*}/\mathrm{M}_{\odot}) = [10.3, 11.7]$ and $[10.4, 11.1]$ for Blue Jay QG and EXCELS, respectively). Higher mass galaxies have deeper potential wells, meaning they are able to hold onto gas and dust more easily than objects at lower mass. Significant dust shielding is needed to see excess Na$\,$\textsc{d} absorption, due to its low ionization potential \citep[5.1$\,$eV; e.g.,][]{chen_absorption-line_2010, roberts-borsani_prevalence_2019}. Redshift effects could also play a role, as galaxies at lower redshift (i.e., the Blue Jay QG sample) have longer dust removal time-scales than high-redshift galaxies \citep[e.g.,][]{lesniewska_dust_2025}. Since dust attenuation and the strength of Na\,\textsc{d} absorption are correlated \citep[e.g.,][]{chen_absorption-line_2010, veilleux_cool_2020, kehoe_aurora_2025}, lower-redshift and higher-mass galaxies may be more likely to exhibit excess Na\,\textsc{d} absorption. Overall, the small sample sizes of both EXCELS and Blue Jay QG, as well as the need for higher-resolution spectroscopy, make drawing any firm conclusions difficult.

\begingroup
\renewcommand{\arraystretch}{1.5}
\begin{table*}
    \centering
        \begin{tabular}{lcccccccccc}
    \hline
    
        EXCELS--ID & &
        \makecell{W$_{\textrm{Na\,\textsc{d}}}$ \\ (\AA)} & &
        \makecell{W$_{\textrm{ISM}}$ \\ (\AA)} & &
        \makecell{$\Delta v$ \\ (km\,s$^{-1}$)} & &
        \makecell{$\sigma_{\mathrm{Na\,\textsc{d}}}$ \\ (km\,s$^{-1}$)} & &
        \makecell{$v_{\mathrm{flow}}$ \\ (km\,s$^{-1}$)} \\
    \hline
    \hline
    
    \multicolumn{11}{c}{Outflow} \\
    
    \hline

    50789 & & 
    5.20 $\pm$ 0.66 & & 
    4.34 $\pm$ 0.62 & & 
    \hphantom{$-$}240$\genfrac{}{}{0pt}{}{+150}{-160}$ & & 
    510$\genfrac{}{}{0pt}{}{+140}{-90}$ & & 
    \hphantom{$-$}1290$\genfrac{}{}{0pt}{}{+320}{-220}$ \\
    
    57000 & & 
    5.59 $\pm$ 0.53 & & 
    4.99 $\pm$ 0.52 & & 
    \hphantom{$-$0}60$\genfrac{}{}{0pt}{}{+40}{-50}$ & & 
    120$\genfrac{}{}{0pt}{}{+60}{-60}$ & & 
    \hphantom{$-$0}310$\genfrac{}{}{0pt}{}{+120}{-100}$ \\
    
    109760 & & 
    7.48 $\pm$ 0.33 & & 
    6.70 $\pm$ 0.41 & & 
    \hphantom{$-$0}90$\genfrac{}{}{0pt}{}{+30}{-30}$ & & 
    190$\genfrac{}{}{0pt}{}{+40}{-50}$ & & 
    \hphantom{$-$0}470$\genfrac{}{}{0pt}{}{+80}{-100}$ \\

    \hline

        \multicolumn{11}{c}{Inflow} \\

    \hline
    
    45981 & & 
    3.55 $\pm$ 0.47 & & 
    2.43 $\pm$ 0.43 & & 
    $-$430$\genfrac{}{}{0pt}{}{+40}{-20}$ & & 
    \hphantom{0}60$\genfrac{}{}{0pt}{}{+60}{-40}$ & & 
    \hphantom{0}$-$550$\genfrac{}{}{0pt}{}{+120}{-80}$ \\
    
    117560 & & 
    3.15 $\pm$ 0.45 & & 
    2.83 $\pm$ 0.48 & & 
    $-$100$\genfrac{}{}{0pt}{}{+70}{-60}$ & & 
    110$\genfrac{}{}{0pt}{}{+80}{-70}$ & & 
    \hphantom{0}$-$320$\genfrac{}{}{0pt}{}{+170}{-140}$ \\

    \hline

        \multicolumn{11}{c}{No residual/excess absorption} \\

    \hline

    34495 & & 
    1.16 $\pm$ 0.37 & & 
    0.44 $\pm$ 0.24 & & 
    -- & & 
    -- & & 
    -- \\
    
    39063 & & 
    1.45 $\pm$ 0.55 & & 
    0.86 $\pm$ 0.45 & & 
    -- & & 
    -- & & 
    -- \\
    
    55410 & & 
    3.12 $\pm$ 0.30 & & 
    1.14 $\pm$ 0.41 & & 
    -- & & 
    -- & & 
    -- \\
    
    65915 & & 
    2.22 $\pm$ 0.47 & & 
    1.23 $\pm$ 0.55 & & 
    -- & & 
    -- & & 
    -- \\
    
    94982 & & 
    0.78 $\pm$ 0.24 & & 
    0.11 $\pm$ 0.08 & & 
    -- & & 
    -- & & 
    -- \\
    
    106260 & & 
    5.49 $\pm$ 1.69 & & 
    5.64 $\pm$ 2.89 & & 
    -- & & 
    -- & & 
    -- \\
    
    113667 & & 
    0.41 $\pm$ 0.52 & & 
    0.44 $\pm$ 0.30 & & 
    -- & & 
    -- & & 
    -- \\
    
    127460 & & 
    0.93 $\pm$ 0.26 & & 
    0.19 $\pm$ 0.15 & & 
    -- & & 
    -- & & 
    -- \\

    \hline
        
    \end{tabular}
    \caption{Outflow properties derived from the best-fitting models to our galaxy sample. Na$\,$\textsc{d} equivalent widths, W$_{\textrm{Na\,\textsc{d}}}$, are calculated following the method outlined in Section \ref{subsub:ppxf}, and the best-fitting model equivalent widths of the residual ISM, W$_{\mathrm{ISM}}$, are calculated following Section \ref{subsub:classification}. Velocity offsets, $\Delta v$, and absorption line widths, $\sigma_{\mathrm{Na\,\textsc{d}}}$, are those of the best-fitting model (see Sections \ref{subsub:model} and \ref{subsub:classification}). Gas flow velocities are calculated using Equation \ref{eq:vflow}.}
    \label{tab:results}
\end{table*}

\endgroup

\subsubsection{Possible Effects of Mergers}\label{subsub:mergers}

One of the main formation pathways for low-$z$ PSBs is thought to be mergers funnelling gas into the centre of the galaxy, triggering a large starburst followed by rapid quenching \citep[e.g.,][]{goto_266_2005, yang_detailed_2008, pawlik_shape_2016, pawlik_origins_2018, wilkinson_merger_2022, leung_spatially_2026}. It is therefore possible that any absorption line offsets could arise from ongoing mergers or galaxy interactions. Fig. \ref{fig:cutouts} shows $2\,^{\prime\prime}\times2\,^{\prime\prime}$ cutout images of our sample. On initial inspection, none of the galaxies show signs of ongoing mergers or interactions, although we note the presence of a possible close companion in the EXCELS--50789 imaging. However, photometric redshift estimations of the companion place it at $z_{\mathrm{phot}}\sim 3.02$, while EXCELS--50789 has $z_{\mathrm{spec}} = 3.99$. This is discussed further in Section \ref{subsub:50789}.

To more quantitatively rule out the presence of ongoing and recent mergers and interactions in our sample, we follow Maltby et al. (in prep.) to investigate three indices commonly used to assess galaxy structural disturbances : residual flux fraction \citep[$RFF$;][]{hoyos_new_2012}, galaxy asymmetry \citep[$A_{\mathrm{gal}}$;][]{conselice_relationship_2003} and the asymmetry of the residual flux once the best-fitting S\'ersic models (see Section \ref{sub:structure}) are removed ($A_{\mathrm{res}}$). We direct the reader to Maltby et al. (in prep.) and \cite{hoyos_new_2012} for full details. We find that the $RFF$ and $A_{\mathrm{gal}}$ values for our galaxies do not indicate any significant level of disturbance. Additionally, only one galaxy, EXCELS--57000, has $A_{\mathrm{res}}$ > 0.6, found by Maltby et al. (in prep.) to indicate potential \textit{late-stage} signatures of a major merger for PSBs with log$_{10}(M_{*}/\mathrm{M}_{\odot}) \geq 10.5$. While one of three galaxies with outflows (EXCELS--57000) shows possible signs of a past merger, all galaxies in the sample fall well into the non-merger region of the $RFF-A_{\mathrm{res}}$ diagnostic diagram presented in \cite{hoyos_new_2012}. Therefore, we conclude that the likelihood of ongoing mergers having a significant impact on our outflow analysis is extremely low.

\subsubsection{EXCELS--50789}\label{subsub:50789}

\begin{figure*}
    \centering
    \includegraphics[width=\textwidth]{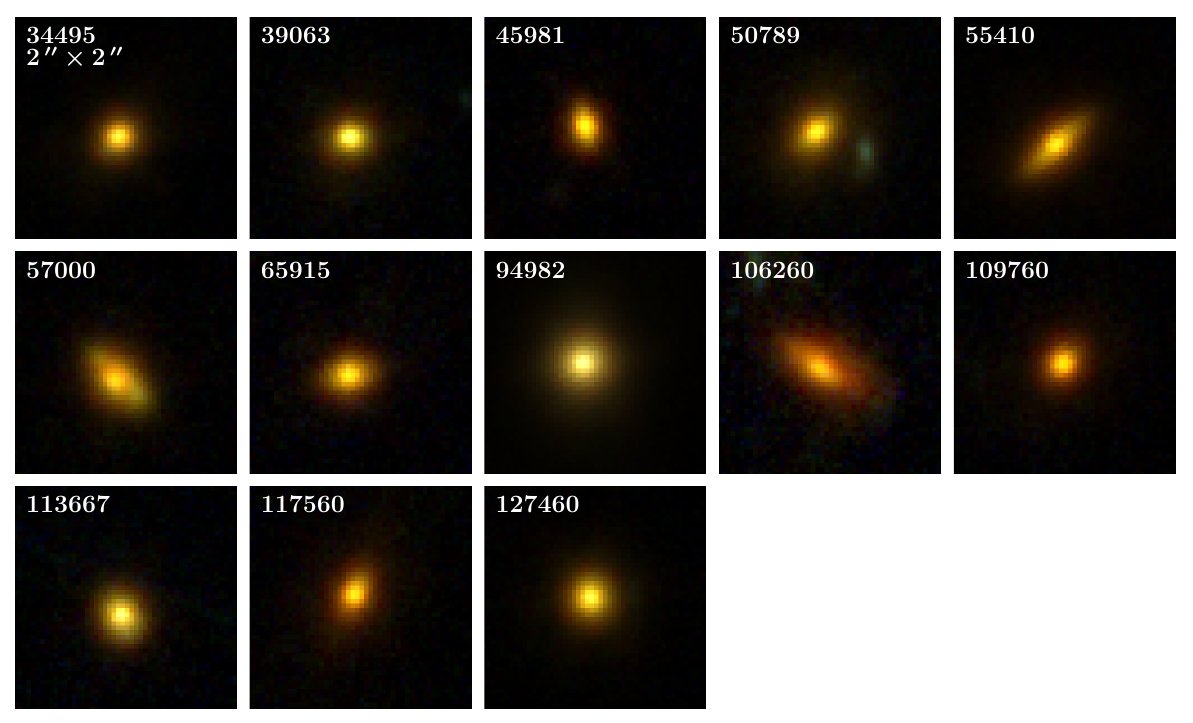}
    \caption{\textit{JWST} PRIMER imaging cutouts ($2\,^{\prime\prime}\times2\,^{\prime\prime}$) for our sample. Galaxy IDs are shown in the top left corner of each cutout. We use the F356W + F444W, F200W + F277W and F090W + F115W filters for the red, green and blue components of the RGB images, respectively. No obvious signs of ongoing or recent mergers are visible for any of our galaxies.}
    \label{fig:cutouts}
\end{figure*}

The galaxy EXCELS--50789 ($z_{\mathrm{spec}} = 3.99$; top-left panel of Fig. \ref{fig:nadflow}) has an extremely broad Na$\,$\textsc{d} profile ($\sigma_{\mathrm{Na\,\textsc{d}}} \approx 500$\,km\,s$^{-1}$). The breadth and positioning of the absorption profile implies inflow as well as outflow, and this extreme width results in a large outflow velocity as derived from Equation \ref{eq:vflow} ($> 1200\,\mathrm{km\,s}^{-1}$, see Table \ref{tab:results}). We note that the possible presence of inflowing gas, as indicated by absorption redward of the Na$\,$\textsc{d} systemic wavelength, will affect the best-fitting $\sigma_{\mathrm{Na\,\textsc{d}}}$ value. However, the resolution of the available spectroscopic data makes separate modelling of the absorption profile via multiple velocity components impractical.

Due to the unusually broad Na\,\textsc{d} profile for EXCELS--50789, we perform additional checks to attempt to rule out alternative explanations not originating from gas flows, such as galaxy rotation and data artefacts. Due to the redshift of EXCELS--50789, the sodium feature is present in both the G235M and G395M data. Both gratings show similarly broad absorption profiles, indicating that this is likely a real feature and not an artefact from the data reduction process. The source is very compact, thus we cannot reliably extract spectra from other pixels in the 2D data products. However, on visual inspection, no absorption lines in the 2D spectrum show signs of galaxy rotation. 

PRIMER imaging of the source (panel 4 of Fig. \ref{fig:cutouts}) indicates that EXCELS--50789 may have a small companion in very close proximity, however this companion was not covered by the slit during observations, and photometric redshift estimations of the companion place it at $z_{\mathrm{phot}} \sim 3.02$. Spectroscopic follow up of the companion redshift is necessary to fully rule out whether EXCELS--50789 is undergoing an interaction. There is also an ALMA source, AS2UDS 0308.0, approximately $3^{\prime\prime}$ from EXCELS--50789 \citep[][]{stach_alma_2018} with a consistent photometric redshift, $z_\mathrm{phot}\simeq4$, which could suggest that EXCELS--50789 exists in a dense environment. 

EXCELS--50789 (otherwise known as ZF--UDS--6496; e.g., \citealt{nanayakkara_population_2024}) was one of the four ultra-massive galaxies discussed in \cite{carnall_jwst_2024}, with a stellar mass of log$_{10}(M_*/\mathrm{M}_\odot)\simeq11$ and an age of $\simeq500$\,Myr at $z=3.99$. Its extreme star-formation history and unusual neutral gas properties make this an important target for further observations, both with JWST at higher spectral resolution, as well as potential future spatially resolved observations with the Extremely Large Telescope.

\subsection{Comparison With Galaxy Properties}\label{section:propertycomparison}

We now examine outflow velocity as a function of time since quenching, along with sources from the literature, as shown in Fig. \ref{fig:tsq}. For completeness, we also include galaxies with non-detections or absorption at systemic velocity, plotted at $v_{\mathrm{flow}} = 0$\,km\,s$^{-1}$. In Section \ref{subsub:props}, we also comment on the lack of correlation seen between outflow velocity and other host galaxy properties.

\subsubsection{Time Since Quenching}\label{subsub:tsq}

\begin{figure*}
    \centering
    \includegraphics[width=0.6\textwidth]{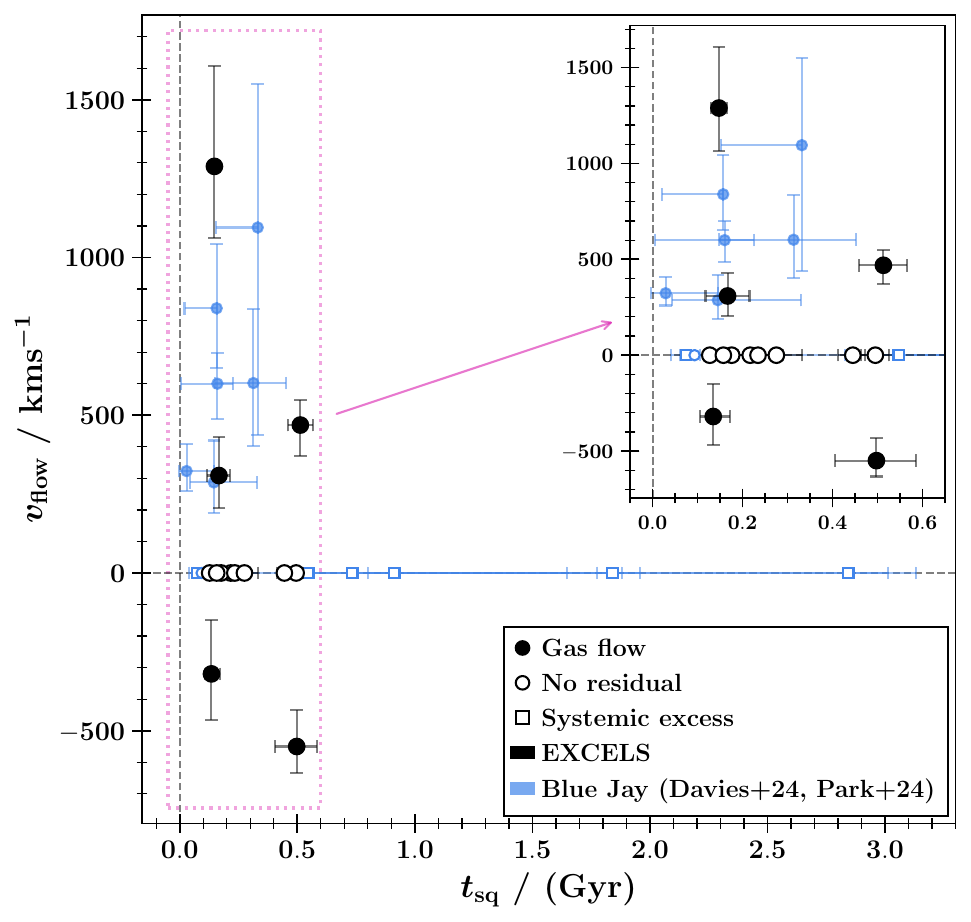}
    \caption{Velocity of gas flow ($v_{\mathrm{flow}}$) versus time since quenching ($t_{\mathrm{sq}}$, see Section \ref{sub:properties}). EXCELS galaxies are shown with black points, and galaxies from the Blue Jay survey (\protect\citetalias{davies_jwst_2024}, see Section \ref{sec:compsamp}) are shown in blue. Galaxies without a detected gas flow are plotted with white centres at $v_{\mathrm{flow}} = 0$\,km\,s$^{-1}$ for completeness. In this work, positive $v_{\mathrm{flow}}$ values indicate an outflow, and negative values indicate an inflow. The dashed grey vertical line represents $t_{\mathrm{sq}} = 0$\,Gyr. The inset axis shows a zoomed in region of the plot, indicated by the magenta dashed rectangle. Gas flows are found exclusively in galaxies with $t_{\mathrm{sq}} \lesssim 600$\,Myr.}
    \label{fig:tsq}
\end{figure*}

In Fig. \ref{fig:tsq}, we compare outflow velocity to the estimated time since quenching, $t_{\mathrm{sq}}$ (Equation \ref{eq:tsq}). It is worth noting that, for quiescent galaxies at these redshifts, the values of $t_{\mathrm{sq}}$ are relatively insensitive to the choice of SFH model used (e.g., see Leung et al. in prep). Estimates of $t_{\mathrm{sq}}$ are calculated for the Blue Jay QG sample using the redshifts of quenching and observation reported in \cite{park_widespread_2024}, which are defined in the same way as for our SFHs.

Fig. \ref{fig:tsq} shows that there are no outflow detections past $t_{\mathrm{sq}} = 0.6$\,Gyr: all the galaxies with outflows in EXCELS and Blue Jay were quenched within the past 600\,Myr. The two objects with signs of inflowing gas have $0.13 \leq t_{\mathrm{sq}} \leq 0.50$\,Gyr. The combined EXCELS and Blue Jay QG sample has a median time since quenching of $\hat{t}_{\mathrm{sq}} = 0.23 \pm 0.10$\,Gyr. Median time since quenching values are $\hat{t}_{\mathrm{sq}} = 0.16 \pm 0.03$\,Gyr for galaxies with detected gas flows (EXCELS and Blue Jay QG), $0.82 \pm 0.51$\,Gyr for galaxies with systemic absorption (Blue Jay QG) and $0.22 \pm 0.06$\,Gyr for galaxies with no excess absorption (EXCELS and Blue Jay QG). \cite{park_widespread_2024} report that, within the Blue Jay QG sample, neutral gas outflows as traced by sodium were found only for galaxies quenched within the past 500$\,$Myr, regardless of the epoch of formation\footnote{\cite{park_widespread_2024} define the epoch of formation of a galaxy as the age of the Universe at which 50 per cent of the observed stellar mass was formed.}. The EXCELS results are consistent with these findings. 

In the local Universe, \citetalias{sun_evolution_2024} found outflow velocities decrease with time elapsed since the end of a starburst (`post-burst age'). While there is no obvious outflow-velocity trend with time since quenching in Fig. \ref{fig:tsq}, our results suggest that, at $z > 1.8$, gas flows are more likely to be detected in galaxies with a quenching event within the past $\sim$ 600\,Myr. We note that the post-burst age parameter used in \citetalias{sun_evolution_2024} traces a slightly different time-scale to $t_{\mathrm{sq}}$ as used in this work, specifically the time since 90 per cent of the stars formed in the most recent burst. 

In contrast, in \cite{taylor_high-velocity_2024}, we found that high-velocity winds ($\Delta v > 1000$\,km\,s$^{-1}$) persisted in PSBs at $1 < z < 1.5$ up to 1\,Gyr after the last burst of star formation. However, \cite{taylor_high-velocity_2024} investigated outflows traced by Mg$\,$\textsc{ii} via a stacking analysis, which has a higher ionization potential (15.0\,eV) than Na$\,$\textsc{d} (5.1\,eV). Mg$\,$\textsc{ii} may therefore trace a different gas phase to Na$\,$\textsc{d}: the latter requires dust shielding to survive \citep[e.g.,][]{chen_absorption-line_2010, roberts-borsani_outflows_2020}, while magnesium may be found in more widespread environments within the ISM and circum-galactic medium (see section 6.3 of \citealt{{kornei_properties_2012}} for a more detailed discussion of how the choice of outflow tracer may affect results). Similarly to \citetalias{sun_evolution_2024}, \cite{taylor_high-velocity_2024} employed a slightly different definition for the evolutionary sequence, using time since \textit{burst} rather than time since quenching.

\subsubsection{Potential Correlations With Other Parameters}\label{subsub:props}

Much of the literature has explored correlations between outflow velocity and intrinsic galaxy properties, such as stellar mass, SFR, SFR surface density and galaxy inclination.
There is still much debate over whether any correlation between outflow velocity and these parameters exists, along with any redshift dependence \citep[e.g.,][]{weiner_ubiquitous_2009, chen_absorption-line_2010, heckman_systematic_2015, sugahara_fast_2019, davis_extending_2023}. The vast majority of studies focus on starburst and normal star-forming galaxies for their analyses, and the absorption lines probed vary across different works, introducing large systematics when comparing reported correlations. 

When investigating these correlations within our sample, we find that there is no clear trend observed between $v_{\mathrm{flow}}$ and any of the host galaxy properties listed above. The lack of correlation holds when we include the literature comparison samples also. This is perhaps not surprising, considering our small sample size and the narrow dynamic range of galaxy properties probed in this work. We have selected galaxies that have low (or fading) SFRs by design, and as such we would likely not expect to see a correlation with SFR or $\Sigma_{\mathrm{SFR}}$.

\section{Outflow Energetics}\label{section:energetics}

Direct causal links between galactic winds and both galaxy quenching and the maintenance of quiescence are still uncertain. However, we can constrain whether an outflow could have contributed to these processes by examining the properties of the wind itself. In this section, we attempt to determine whether current observed levels of star formation and AGN activity in our sample could plausibly drive these winds, by calculating mass, energy and momentum outflow rates. We largely follow the methodology of \citetalias{davies_jwst_2024} in this section.

\subsection{Outflow Properties}\label{sub:mout}

\begingroup
\renewcommand{\arraystretch}{1.5}
\begin{table*}
    \centering
    \begin{tabular}{lcccccccccccccc}
    \hline
    
    EXCELS--ID & & 
    \makecell{$v_{\mathrm{flow}}$ \\ (km\,s$^{-1}$)} & &          \makecell{log$_{10}$(N(Na$\,$\textsc{i})) \\ (cm$^{-2}$)} & &  
    \makecell{log$_{10}$(N(H$\,$\textsc{i})) \\ (cm$^{-2}$)} & & 
    \makecell{log$_{10}$($\dot{M}_{\mathrm{out}}$) \\ (M$_{\odot}$\,yr$^{-1}$)} & & 
    \makecell{log$_{10}$($\dot{E}_{\mathrm{out}}$) \\ (erg\,s$^{-1}$)} & & 
    \makecell{log$_{10}$($\dot{p}_{\mathrm{out}}$) \\ (dyne)} & & 
    log$_{10}$($\eta$) \\
    \hline
    \hline

50789 & & 
1290$\genfrac{}{}{0pt}{}{+320}{-220}$ & & 
14.8$\genfrac{}{}{0pt}{}{+0.2}{-0.4}$ & & 
22.4$\genfrac{}{}{0pt}{}{+0.2}{-0.4}$ & & 
2.3$\genfrac{}{}{0pt}{}{+0.2}{-0.3}$ & & 
44.0$\genfrac{}{}{0pt}{}{+0.4}{-0.5}$ & & 
36.2$\genfrac{}{}{0pt}{}{+0.3}{-0.4}$ & & 
> 3.1\\

57000 & & 
\hphantom{0}310$\genfrac{}{}{0pt}{}{+120}{-100}$ & & 
13.9$\genfrac{}{}{0pt}{}{+0.3}{-0.4}$ & & 
21.6$\genfrac{}{}{0pt}{}{+0.3}{-0.4}$ & & 
1.5$\genfrac{}{}{0pt}{}{+0.2}{-0.3}$ & & 
42.0$\genfrac{}{}{0pt}{}{+0.5}{-0.6}$ & & 
34.8$\genfrac{}{}{0pt}{}{+0.4}{-0.5}$ & & 
> 1.6\\

109760 & & 
\hphantom{0}470$\genfrac{}{}{0pt}{}{+80}{-100}$ & & 
13.8$\genfrac{}{}{0pt}{}{+0.2}{-0.2}$ & & 
21.4$\genfrac{}{}{0pt}{}{+0.2}{-0.2}$ & & 
1.7$\genfrac{}{}{0pt}{}{+0.1}{-0.1}$ & & 
42.5$\genfrac{}{}{0pt}{}{+0.2}{-0.3}$ & & 
35.2$\genfrac{}{}{0pt}{}{+0.2}{-0.2}$ & & 
> 2.6\\
    \hline

\end{tabular}
    \caption{Outflow properties and energetics for the three galaxies in our sample with detected outflows, derived following the process described in Section \ref{sub:mout}. Lower limits on the mass loading factors, $\eta$, are calculated using the 16th percentile of $\dot{M}_{\mathrm{out}}$ from Bayesian analysis, and SFR upper limits from Table~\ref{tab:props}. Systematic uncertainties on the values of the metal ionization fraction, the opening angle of the wind and the outflow radius (discussed in Section \ref{sub:mout}) are not included in the quoted errors.}
    \label{tab:disc}
\end{table*}
\endgroup

From our best-fitting line profiles, we can roughly estimate the mass outflow rates of the winds, $\dot{M}_{\mathrm{out}}$. To do this, we first estimate the column densities of Na$\,$\textsc{d}. Following \cite{draine_physics_2011}, the column density of sodium, N(Na$\,$\textsc{i}), can be derived using:
\begin{multline}
    \mathrm{N(Na~\textsc{i})} = 10^{13}\, \mathrm{cm}^{-2}\bigg(\frac{\hphantom{0}\tau_{0}\hphantom{0}}{0.7580} \bigg)\left(\frac{0.4164}{f_{\ell u}}\right)\\
    \times\left(\frac{1215\text{\AA}}{\lambda_{\ell u}}\right)\left(\frac{b}{10\,\mathrm{km\,s}^{-1}}\right).
\end{multline}
Here, $f_{\ell u} = 0.32$ and $\lambda_{\ell u} = 5897$\,\AA\ are the oscillator strength and the rest-frame wavelength of the transition, respectively, $\tau_{0}$ is the optical depth derived from our model fitting (Section \ref{subsub:model}), and $b = \sqrt{2}\sigma_{\mathrm{Na\,\textsc{d}}}$ is the Doppler parameter, where $\sigma_{\mathrm{Na\,\textsc{d}}}$ is the line width given by the best-fitting models. We note that if the absorption lines are saturated, the estimated column density would be a lower limit, thus increasing the true mass outflow rate.

The column density of hydrogen, N(H$\,$\textsc{i}), can then be estimated from N(Na$\,$\textsc{i}), using:
\begin{equation}
    \mathrm{N(H~\textsc{i})} = \frac{\mathrm{N(Na~\textsc{i})}}{\chi\mathrm{(Na~\textsc{i})} \, 10^{\,a + d}},
\end{equation}
where $\chi$(Na$\,$\textsc{i}) is the metal ionization fraction, $a$ = log(Na$\,$\textsc{i}/H$\,$\textsc{i}) is the metal abundance relative to H, and $d$ traces the level of dust depletion \citep{rupke_outflows_2005, veilleux_cool_2020}. We adopt the values used in \citetalias{davies_jwst_2024}, of $\chi$(Na$\,$\textsc{I}) = 0.1 \citep[][]{stokes_interstellar_1978, stocke_new_1991}, $a$(Na$\,$\textsc{i}) = $-$5.69 and $d$(Na$\,$\textsc{i}) = $-$0.95 \citep[][]{savage_interstellar_1996}. Typical uncertainties on $a$(Na$\,$\textsc{i}) and $d$(Na$\,$\textsc{i}) are $\sim 0.04$ and 0.1\, dex, respectively \citep[see][]{savage_interstellar_1996}. Recent work by \cite{moretti_empirical_2025} found good agreement between the local N(Na$\,$\textsc{i})--N(H$\,$\textsc{i}) conversion, and that derived empirically for a massive quiescent galaxy at $z = 2.4$. However, as discussed in \citetalias{davies_jwst_2024}, the metal ionization fraction of 10 per cent is based on stars within the Milky Way and an extragalactic H\,\textsc{i} cloud, and higher neutral fractions have been found in local AGN winds \citep[$\chi$(Na$\,$\textsc{I}) = 0.05;][]{baron_multiphase_2020}. A lower ionization fraction would increase the hydrogen column density (and, therefore, mass outflow rate) estimates. 

The mass outflow rate for a simple shell geometry can be written as:
\begin{multline}
    \dot{M}_{\mathrm{out}} = 11.45\,\mathrm{M}_{\odot}\mathrm{\,yr}^{-1}\left(C_{\Omega}\frac{C_{f}}{0.4}\right)\left(\frac{\mathrm{N(H~\textsc{i})}}{10^{21}\mathrm{\,cm}^{-2}}\right) \\
    \times \left(\frac{r_{\mathrm{out}}}{1\mathrm{\,kpc}}\right)\left(\frac{v_{\mathrm{flow}}}{200 \mathrm{\,km\,s}^{-1}}\right).
\end{multline}
Following \citetalias{davies_jwst_2024}, we set the large-scale covering factor due to the wind opening angle as $C_{\Omega}$ = 0.5, and the extent of the outflow, $r_{\mathrm{out}} = 1$\,kpc. We derive uncertainties in our N(Na$\,$\textsc{i}), N(H$\,$\textsc{i}) and $\dot{M}_{\mathrm{out}}$ values using the full posterior distributions estimated from our Bayesian fitting (Section \ref{sub:fitting}). The main uncertainties in our $\dot{M}_{\mathrm{out}}$ estimates originate from the estimates of N(Na$\,$\textsc{i}) and N(H$\,$\textsc{i}), as discussed above, as well as the unknown value of $C_{\Omega}$: as discussed in \citetalias{davies_jwst_2024} and \cite{sun_extreme_2026}, $C_{\Omega} = 0.5$ assumes that the winds cover 50 per cent of the solid sphere. \citetalias{davies_jwst_2024} argue $C_{\Omega}$ is highly unlikely to be less than 0.25 and therefore we adopt a conservative systematic uncertainty on $C_{\Omega}$ of a factor of two. The $r_{\mathrm{out}}$ parameter is also highly uncertain: for two of our galaxies (EXCELS--50789 and EXCELS--57000), an outflow extent of $r_{\mathrm{out}} = 1$\,kpc is smaller than the measured $r_{\mathrm{eff}}$ (see Table \ref{tab:props}), and studies such as \cite{deugenio_fast-rotator_2024} have found outflows extending out to $\sim 3$\, kpc. Additionally, excess cool gas absorption (traced by Mg\,\textsc{ii}) has been found out to several 100\,kpc in the circum-galactic medium surrounding PSBs \citep[][]{harvey_cool_2025}. Thus, our $r_{\mathrm{out}}$ estimate is likely very conservative, setting a lower limit on our mass outflow rates. The estimates of column density and mass outflow rate are presented in Table \ref{tab:disc}, and we note that we have not included the systematics discussed above in the quoted (random) errors.

Given the above assumptions, we find mass outflow rates of $\dot{M}_{\mathrm{out}} \simeq 20-200\,\mathrm{M}_{\odot}\,\mathrm{yr}^{-1}$, over two orders of magnitude higher than the current SFRs of our sample, implying that current levels of star formation are unlikely to drive the observed winds. We explore this point further in Section \ref{sub:sfwinds}. The large $\dot{M}_{\mathrm{out}}$ values as compared to the galaxy SFRs also indicates that the observed outflows are likely capable of playing a significant role in keeping these galaxies quenched, regardless of whether winds are explicitly linked to the initial quenching event or not. To test this further, we compare $v_{\mathrm{flow}}$ to an estimate of the galaxy escape velocity, $v_{\mathrm{esc}}$, using:
\begin{equation}
    v_{\mathrm{esc}} = \sqrt{\frac{2GM_{*}}{r_{\mathrm{eff}}}},
\end{equation}
where \textit{G} is the gravitational constant. We use the galaxy stellar mass to calculate $v_{\mathrm{esc}}$, as high-redshift quiescent galaxies are likely to be stellar dominated within $r_{\mathrm{eff}}$ \citep[within a factor of $\sim 2$; e.g.,][]{de_graaff_ionised_2024, slob_fast_2025, cheng_bottom-heavy_2026}. We find $v_{\mathrm{flow}} \approx 1.5v_{\mathrm{esc}}$ for EXCELS--50789, indicating that at least a fraction of the outflowing gas will escape the galaxy halo, reducing any remaining cold gas supply available for star formation. The outflow velocities for EXCELS--57000 and EXCELS--109760 are lower than our escape velocity estimates (though of the same order of magnitude), and in this case the gas may fall back onto the galaxy as fountains \citep[e.g.,][]{fox_gas_2017, li_fountain-driven_2023}. However, the fate of the inflowing gas remains hard to constrain \citep[see, e.g.,][for further discussion]{bevacqua_feeding_2025}.

The energetics of a galactic wind may help to uncover the driving force behind the outflow. Following \citetalias{davies_jwst_2024}, the energy and momentum rates of an outflow ($\dot{E}_{\mathrm{out}}$ and $\dot{p}_{\mathrm{out}}$, respectively) can be estimated simply, using:
\begin{align}
        \dot{E}_{\mathrm{out}} &= \frac{1}{2}\,\dot{M}_{\mathrm{out}}\,v_{\mathrm{flow}}^{2}\,, \\
        \dot{p}_{\mathrm{out}} &= \dot{M}_{\mathrm{out}}\,v_{\mathrm{flow}}.
\end{align}
Computed energy and momentum outflow rates are also presented in Table \ref{tab:disc}. We compare the $\dot{E}_{\mathrm{out}}$ and $\dot{p}_{\mathrm{out}}$ values derived from our spectroscopic analysis to those expected from theoretical predictions for star formation (i.e., supernovae) and AGN driven winds in Sections \ref{sub:sfwinds} and \ref{sub:agnwinds}.

\subsection{Star Formation Driven Winds}\label{sub:sfwinds}

\begin{figure*}
    \centering
    \includegraphics[width=\textwidth]{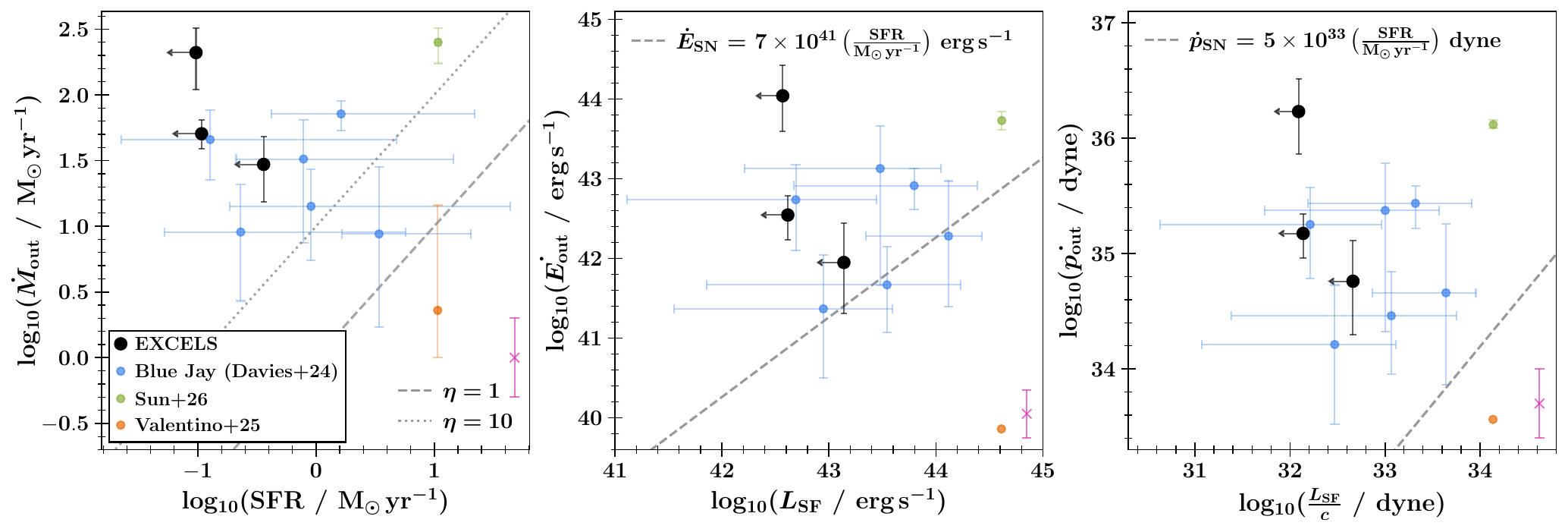}
    \caption{\textit{Left}: Current star-formation rates versus mass outflow rates of the galaxies in our sample. EXCELS galaxies (black points) are plotted at the upper limit of SFR (as given in Table \ref{tab:props}). Galaxies from recent literature are also shown \protect\citep[\citetalias{davies_jwst_2024};][see Section \ref{sec:compsamp}]{sun_extreme_2026, valentino_gas_2025}. The magenta error bar denotes the systematic uncertainty on estimated mass outflow rates (and thus energy and momentum rates) due to the assumption made for the wind opening angle covering factor (see Section \ref{sub:mout}). Mass loading factors, $\eta = \dot{M}_{\mathrm{out}}/\mathrm{SFR}$, of $\eta = 1$ and $\eta = 10$ (the approximate maximum found for star-forming galaxies at $z \sim 3.5$; see Section \ref{sub:sfwinds}) are represented by dashed and dotted lines respectively. The mass loading factors implied for our galaxies are an order of magnitude higher than expected for star-forming galaxies at similar redshifts. \textit{Centre}: Energy outflow rates for galaxies with detected outflows compared to expected values for supernovae driven winds (Equation \ref{eq:sfe}). \textit{Right:} Momentum outflow rates for galaxies with detected outflows compared to expected values for supernovae driven winds (Equation \ref{eq:sfp}). The energy and momentum outflow rates expected from current levels of star formation are insufficient to explain those derived from our absorption line modelling. These results strongly suggest that the current levels of star formation in these galaxies could not plausibly drive the observed winds. }
    \label{fig:sf}
\end{figure*}

As discussed in the previous section, the mass outflow rates estimated from our best-fitting absorption line models are over two orders of magnitude larger than current star-formation rates in our sample. If we begin by assuming that the outflows we observe are driven by star formation, we can compare the SFR of each galaxy to the mass outflow rate, in order to estimate how efficient such star-formation driven outflows would need to be. The ratio $\eta = \dot{M}_{\mathrm{out}}/\mathrm{SFR}$ is known as the mass loading factor, where $\eta > 1$ indicates outflows are removing gas faster than the current rate of star formation. 

We show the calculated mass loading factors for our EXCELS galaxies with outflows, assuming the winds are powered by star formation, in the left panel of Fig. \ref{fig:sf}. We include dashed and dotted lines representing $\eta = 1$ and 10 respectively. A mass loading factor of $\eta = 10$ is roughly the maximum expected for $\mathrm{log}_{10}(M_{*}/\mathrm{M}_{\odot}) \sim 10.75$ star-forming galaxies at $z \sim 3.5$ (e.g., in \textsc{eagle},  \citeauthor{mitchell_galactic_2020} \citeyear{mitchell_galactic_2020}, and using chemical evolution models, \citeauthor{stanton_nirvandels_2024} \citeyear{stanton_nirvandels_2024}). As discussed in \cite{stanton_nirvandels_2024}, mass loading factors for the ionized gas phase in star-forming galaxies derived using UV absorption lines \citep[e.g.,][]{chisholm_mass_2017} are an order of magnitude lower ($\eta \sim 1$) than those derived using chemical abundance models. Even lower mass loading factors derived from ionized gas ($\eta < 1$) have been reported for galaxies in the local Universe \citep[e.g.,][]{marasco_shaken_2023}, and at cosmic noon \citep[e.g.,][]{concas_being_2022}. 

As can be seen from Fig. \ref{fig:sf}, the mass loading factors required to explain our observed outflows via star formation alone are $\eta\gtrsim100$ (even under the conservative assumptions we have made for e.g., ionization state, covering fraction, line saturation). This is much higher than the $\eta\simeq10$ observed for star-forming galaxies, strongly suggesting that the outflows we observe cannot be driven by star formation alone.

To compare our outflow energy and momentum calculations from Section \ref{sub:mout} with predicted energy and momentum injection rates from supernovae, we use the relations from \cite{veilleux_galactic_2005}:
\begin{align}
        \dot{E}_{\mathrm{SN}} &= 7 \times 10^{41}  \bigg(\frac{\mathrm{SFR}}{\mathrm{M}_{\odot}\,\mathrm{yr}^{-1}}\bigg) \,\,\mathrm{erg\,s}^{-1},\label{eq:sfe} \\
        \dot{p}_{\mathrm{SN}} &= 5 \times 10^{33} \bigg(\frac{\mathrm{SFR}}{\mathrm{M}_{\odot}\,\mathrm{yr}^{-1}}\bigg) \,\,\mathrm{dyne}\;\label{eq:sfp}.
\end{align}
The above relations are based on solar metallicity Starburst99 models \citep{leitherer_starburst99_1999}. To compare with recent works \citep[e.g., \citetalias{davies_jwst_2024};][]{ sun_extreme_2026}, we estimate the luminosity of star formation in our galaxies using our \textsc{Bagpipes} SFR upper limits (see Table \ref{tab:props}), $L_{\mathrm{SF}} \approx 10^{10}\,L_{\odot} (\mathrm{SFR}/\mathrm{M}_{\odot}\,\mathrm{yr}^{-1})$. As demonstrated by \cite{stevenson_primer_2025} and \cite{skarbinski_jwst_2025}, none of the galaxies in our sample show emission line ratios consistent with star formation (see Section \ref{sub:drivers} for more detail). Thus, we avoid using H$\alpha$ derived SFRs, as we cannot confidently attribute the observed H$\alpha$ flux to pure star formation.

We compare our model-derived momentum and energy rates to those predicted from supernovae in the centre and right-hand panels of Fig. \ref{fig:sf}. In accordance with our conclusions based on mass-outflow rates, our inferred outflow energy and momentum rates of the observed winds are too high for current star formation to be a plausible driver. Supernovae driven winds have been constrained to have a longevity of $\lesssim 25$\,Myr \citep[e.g.,][]{mcquinn_galactic_2018}, and the time since quenching values for our sample (see Table \ref{tab:props}) are all $>5$ times higher than this. This indicates that previous episodes of star formation are also unlikely to be the cause of these winds.

\subsection{AGN Driven Winds}\label{sub:agnwinds}

\begin{figure*}
    \centering
    \includegraphics[width=0.98\textwidth]{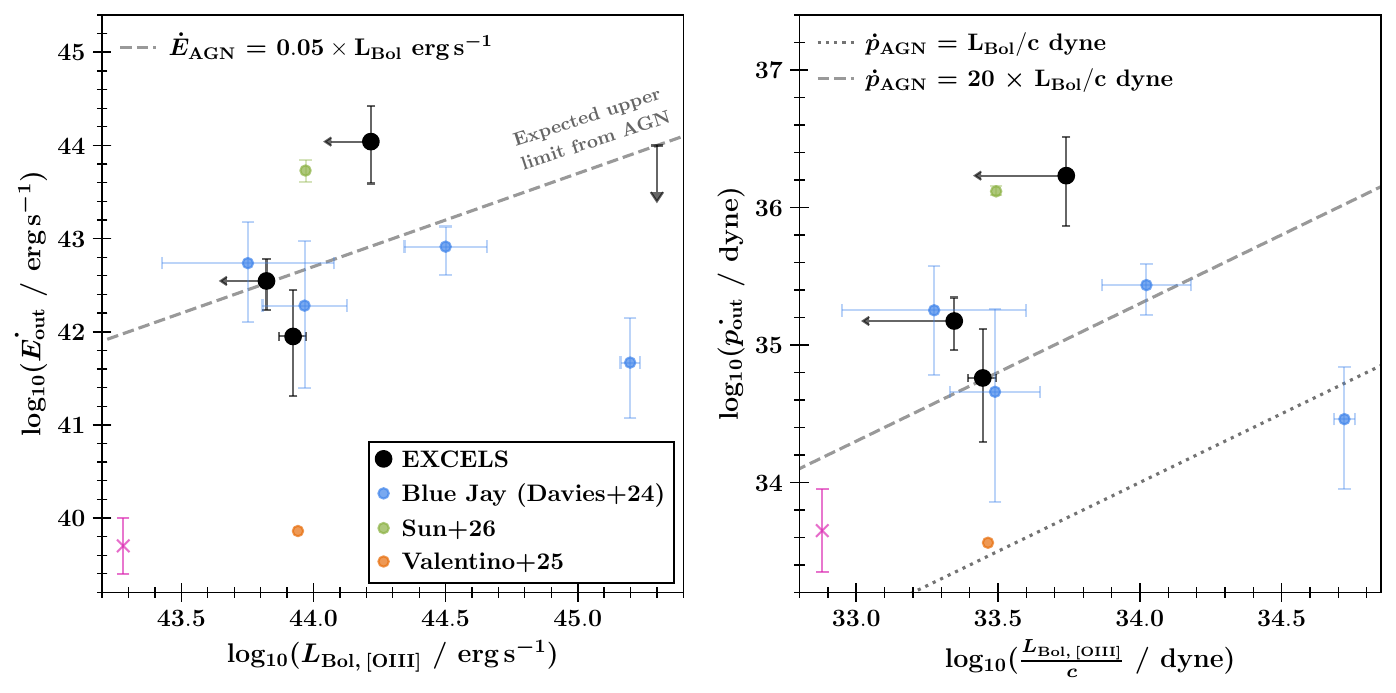}
    \caption{\textit{Left}: Energy outflow rates for galaxies with detected outflows compared to expected values for AGN driven winds (Equation \ref{eq:agne}). Here, $L_{\mathrm{Bol,\, [O~\textsc{iii}]}}$ is an estimate for the bolometric AGN power calculated from the [O~\textsc{iii}] line flux. Two of the three  EXCELS galaxies (black points) are plotted at the $3\sigma$ upper limits on $L_{\mathrm{Bol,\, [O~\textsc{iii}]}}$. Galaxies from recent literature are also shown \protect\citep[\citetalias{davies_jwst_2024};][see Section \ref{sec:compsamp}]{sun_extreme_2026, valentino_gas_2025}. The magenta error bar denotes the systematic uncertainty on the estimated mass outflow rates (and thus energy and momentum rates) due to the assumption made for the wind opening angle covering factor (see Section \ref{sub:mout}). \textit{Right}: Momentum outflow rates for galaxies with detected outflows compared to expected values for AGN driven winds (Equation \ref{eq:agnp}). One galaxy has outflow energetics that could potentially be explained by current AGN activity, while the others have anomalously high energy and/or momentum outflow rates compared to theoretical predictions, suggesting that energy and momentum were injected into the outflows by recent, more-luminous AGN activity, which has since faded.}
    \label{fig:agn}
\end{figure*}

As current levels of star formation are insufficient to explain the observed winds, we then turn to AGN activity as a potential driver. We use the expected energy relation from the energy-conserving scenario of \cite{king_powerful_2015}:
\begin{equation}
        \dot{E}_{\mathrm{AGN}} = 0.05 \times L_{\mathrm{Bol}}\,,\label{eq:agne}
\end{equation}
where $L_{\mathrm{Bol}}$ is the AGN bolometric luminosity. Equation \ref{eq:agne} is derived assuming that all the mechanical luminosity ($L_{\mathrm{BH}} \sim (L_{\mathrm{AGN}}/c)(v/2)$) of an initial, small scale $v \sim 0.1c$ accretion disk wind is transferred to the surrounding ISM. The momentum of the large-scale ISM outflow is then:
\begin{equation}
    \dot{p}_{\mathrm{AGN}} = L_{\mathrm{Bol}} / c \label{eq:agnp}\,,
\end{equation}
where $c$ is the speed of light.

As discussed in \citetalias{davies_jwst_2024}, AGN momentum outflow rates have been observed to be boosted by a factor of $2-20$ in energy-conserving outflows \citep[e.g.,][]{faucher-giguere_physical_2012, cicone_massive_2014}, matching predictions from theoretical models \citep[e.g.,][]{faucher-giguere_physics_2012, king_powerful_2015}. We therefore follow \citetalias{davies_jwst_2024} and additionally compare our estimated momentum rates to $\dot{p}_{\mathrm{AGN}} = 20\,\times \,L_{\mathrm{Bol}} /c$. We estimate $L_{\mathrm{Bol}}$ for our galaxies using the [O$\,$\textsc{iii}] bolometric correction \citep[e.g.,][]{kauffmann_feast_2009,netzer_accretion_2009}, $L_{\mathrm{Bol,\, [O~\textsc{iii}]}} = 600\,L_{\mathrm{[O~\textsc{iii}]}}$. Given the results of \cite{stevenson_primer_2025} and \cite{skarbinski_jwst_2025}, it is likely that any [O$\,$\textsc{iii}] emission in our sample is due to AGN. We note that two of the three galaxies in our outflow sample did not have detectable [O$\,$\textsc{iii}] emission, and therefore their $L_{\mathrm{Bol,\, [O~\textsc{iii}]}}$ values are strictly upper limits, estimated using the $3\sigma$ upper limit of the flux measurement (see \citealt{stevenson_primer_2025} for further details). $L_{\mathrm{Bol}}$ estimates derived from [O$\,$\textsc{iii}] fluxes are additionally subject to systematic uncertainties pertaining to dust obscuration and the physical state of the AGN \citep[e.g.,][]{netzer_bolometric_2019}. Here, we do not attempt to dust-correct our [O$\,$\textsc{iii}] fluxes and limits (consistent with the literature comparison samples), both because our quiescent galaxies are not expected to contain significant dust, and because dust attenuation is extremely difficult to reliably constrain from spectral fitting (e.g., \citealt{pacifici_art_2023}).

Fig. \ref{fig:agn} shows the estimated $\dot{E}_{\mathrm{AGN}}$ (left) and $\dot{p}_{\mathrm{AGN}}$ (right) values for our sample, versus those derived from our the best-fitting outflow model. Focusing first on the energy rates, we find that two of our three galaxies have $\dot{E}_{\mathrm{out}}$ plausibly explained by current AGN activity. When comparing to AGN momentum rates, we find that only one galaxy may have $L_{\mathrm{Bol,\, [O~\textsc{iii}]}}$ sufficient to explain the derived $\dot{p}_{\mathrm{out}}$ value. This object is EXCELS--57000, which has line ratios consistent with a current AGN (detailed further in Section \ref{sub:drivers}). For EXCELS--57000, we find $\mathrm{log}_{10}(\dot{p}_{\mathrm{out}}) \sim 34.5$ and $\mathrm{log}_{10}(\dot{p}_{\mathrm{AGN}}) \sim 33.4$, implying an approximate momentum boost of a factor of 10. However, it should be noted that the AGN coupling efficiency of 5 per cent assumed in energy-conserving outflows (Equation \ref{eq:agne}) is believed to operate only when close to the Eddington limit \citep[e.g.,][]{cicone_massive_2014}. Most of the AGN present in our sample are likely far below Eddington, so the coupling efficiency is expected to be lower. Therefore, a more natural explanation is that the outflows were driven by earlier, more luminous AGN activity, for which the injected momentum rates can still be seen in outflows despite the AGN fading away.

Similar results were found for JADES-GS-206183 in \citeauthor{sun_extreme_2026} (\citeyear{sun_extreme_2026}; green circle in Figs \ref{fig:sf} and \ref{fig:agn}), who determine the wind is a fossil outflow driven by recent episodic AGN activity. \cite{fluetsch_cold_2018} found around 10\,--\,20 per cent of their sample of local star-forming and AGN host galaxies displayed signs of fossil outflows. These objects also displayed anomalously high energy and momentum rates compared to the expected values derived from $L_{\mathrm{Bol}}$ and SFR. \cite{zubovas_determining_2022} predict that fossil outflows should outnumber those driven by \textit{current} AGN activity by 1.6:1, with approximately 20 per cent of all galaxies showing evidence for AGN related outflows (both fossil and current). Our results are consistent with these findings.

\section{A simple model for fossil outflows and their time-scales}\label{sec:time}

In Section \ref{section:energetics} we have demonstrated that residual star formation cannot power the outflows we observe in our sample, and that current levels of AGN activity are unlikely to be sufficient in all cases. In this section, we therefore explore episodic AGN activity post quenching as a potential driver of the observed winds, under the assumption that, in some cases, the AGN fades whilst the outflow is still visible. We construct a simple `outflow cycle' model based on our observations and estimate the timescales for each phase, making multiple consistency checks using different lines of evidence.

\begin{figure*}
    \centering
    \includegraphics[width=0.7\textwidth]{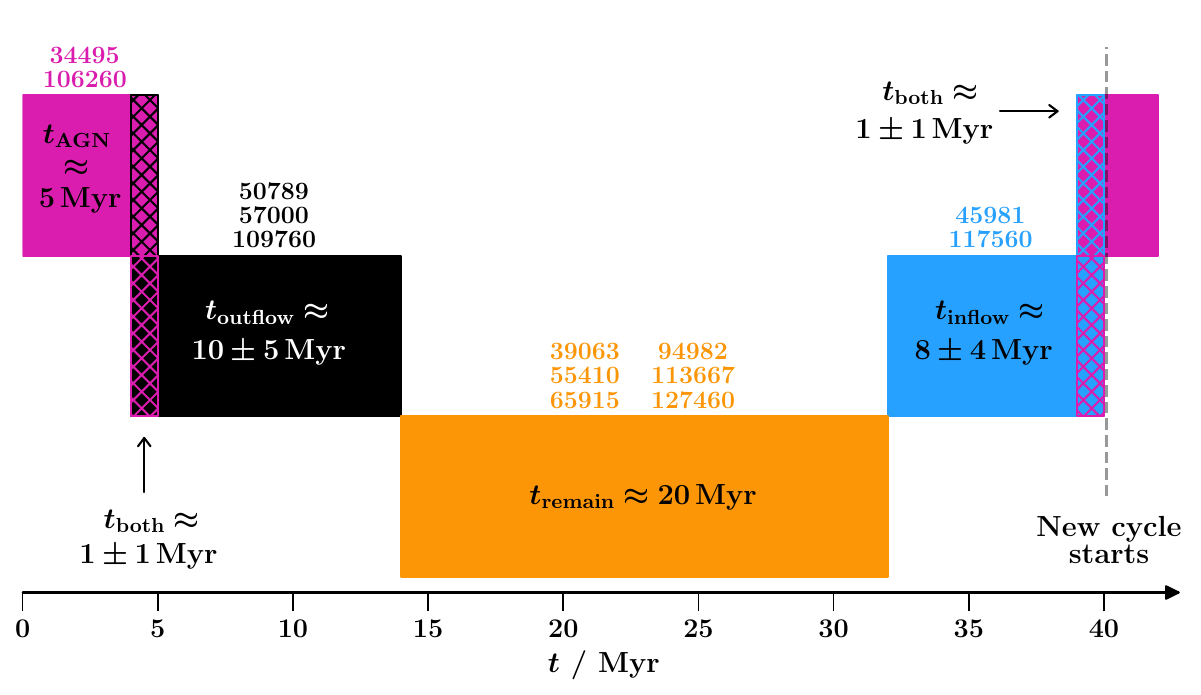}
    \caption{Illustration of a possible outflow cycle for $z\simeq3$ quiescent galaxies, as determined from our observations and the \textsc{eagle} simulation (see Section~\ref{sec:time}), with $t_{\mathrm{cycle}} \sim 40$\,Myr. Each EXCELS object is denoted above the cycle phase in which it is observed. Outflows are launched by energetic AGN (magenta), detectable in X--ray imaging for $t_{\mathrm{AGN}} \sim 5$\,Myr on average (Section \ref{ref:xray}). The probability of observing an active AGN and an outflow simultaneously in our sample corresponds to an overlap time of $t_{\mathrm{both}}\approx1\pm1$\,Myr (hatched black and magenta), after which the AGN fades rapidly, and fossil outflows are seen (black). These winds retain the energy and momentum rates injected during the active AGN phase, thus leading to anomalously high energetics when compared to current levels of AGN activity (see Section \ref{section:energetics}). Fossil outflows persist for $\simeq9\pm5$\,Myr after the AGN has switched off. The outflows eventually dissipate, and galaxies are observed with no  AGN activity or gas flows for roughly $t_{\mathrm{remain}}\sim 20$\,Myr. After this time, the galaxy begins reaccreting gas from the surrounding halo, which may then fuel another energetic AGN episode, and the cycle begins again.}
    \label{fig:cycle}
\end{figure*}

We begin by assuming that all galaxies in our sample undergo an outflow cycle post quenching. This consists of the AGN switching on, and after a period beginning to drive a visible outflow. The AGN then switches off, after which the outflow persists for a period, then also fades. This is followed by a period of no activity, after which gas begins to inflow again, eventually re-igniting AGN activity. A schematic diagram is presented in Fig. \ref{fig:cycle}.

To guide our discussion, in Section \ref{sub:eagle} we first investigate AGN accretion in simulated $z=3$ quiescent galaxies from the \textsc{eagle} simulation  \citep{schaye_eagle_2015, crain_eagle_2015}.
In Section \ref{sub:drivers} we examine various AGN activity indicators for galaxies in our sample and compute Eddington ratios, then compare the probability of observing various AGN indicators in our sample to the time spent above a given Eddington ratio in \textsc{eagle}. In Section \ref{sub:prob}, we perform a simulation-based inference analysis to compute the probabilities of observing AGN, outflows and inflows in our sample, including the overlaps between these categories. We draw these elements together to form a proposed cycle timeline in Section \ref{sub:timeline}.

\begin{figure}
    \centering
    \includegraphics[width=\columnwidth]{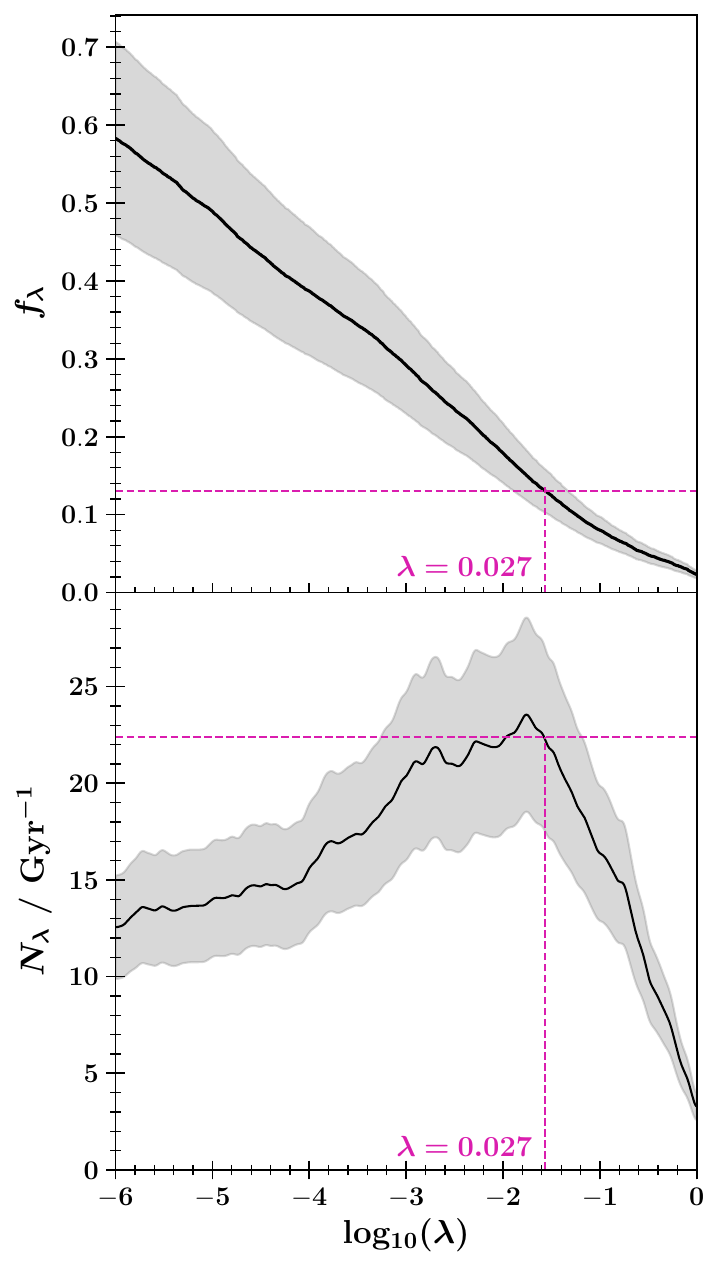}
    \caption{\textit{Top}: Fraction of time post quenching spent above a given Eddington ratio ($f_{\lambda}$) versus Eddington ratio ($\lambda$) for the average quenched $z = 3$ galaxy in the \textsc{eagle} simulation. Shaded grey regions correspond to the standard error on the mean. Magenta dashed lines denote $\lambda = 0.027$, the Eddington ratio above which a galaxy would be X--ray detected in the \textit{Chandra} UDS imaging (see Sections \ref{sub:eagle} and \ref{ref:xray}). \textit{Bottom}: Number of episodes post quenching that the average quenched $z = 3$ \textsc{eagle} galaxy has above a given Eddington ratio per billion years spent quenched, $N_{\lambda}$.}
    \label{fig:eagle}
\end{figure}

\subsection{AGN in Simulated $\mathbf{z=3}$ Quiescent Galaxies from EAGLE}\label{sub:eagle}

To gain an indicative understanding of AGN activity in our galaxies, we make use of the  \textsc{eagle} simulation \citep{schaye_eagle_2015, crain_eagle_2015}. We select galaxies from the $z = 3$ snapshot with a stellar mass of $\mathrm{log}_{10}(M_{*}/\mathrm{M}_{\odot}) > 10$ and a specific star-formation rate of sSFR $< 0.2/t_{\mathrm{H}}(z)$. This stellar mass threshold is lower than the $\mathrm{log}_{10}(M_{*}/\mathrm{M}_{\odot}) = 10.4$ mass-completeness limit for our sample (see \citealt{stevenson_primer_2025}, Leung et al. in prep), however this choice was necessary to provide a sufficiently sized sample of \textsc{eagle} galaxies. We additionally remove objects for which the black hole has not reached a mass $M_{\mathrm{BH}} \geq 10^7$\,M$_{\odot}$, to ensure the black holes are no longer seeds. This results in a sample of 22 simulated galaxies at $z = 3$.

Each galaxy has a $t_{95}$ value, defined as the time at which the object had formed 95 per cent of its stellar mass. We determine the redshift of quenching, $z_{95}$, as the corresponding redshift for which the age of the Universe matches $t_{95}$.

We then investigate the accretion history of the black holes in these simulated galaxies post quenching. For each \textsc{eagle} galaxy, we tabulate the Eddington ratio, $\lambda$ in 1 Myr intervals, then calculate the fraction of time between $z=3$ and $z_{95}$ spent above a given Eddington ratio, $t_{\lambda}$, as well as the average number of episodes a galaxy has above a given Eddington ratio per billion years spent quenched, $N_{\lambda}$. This information is displayed in Fig. \ref{fig:eagle}. It can be seen that significant AGN activity persists in these simulated galaxies after quenching, with $\sim20$ per cent of the time spent above $\lambda=0.01$ (often used as a canonical threshold for AGN activity, e.g., \citealt{schulze_low_2010, shankar_accretion-driven_2013}). AGN activity is also highly variable, with $\simeq15-25$ cycles per Gyr. The distribution of accretion rates shown in Fig. \ref{fig:eagle} is in good qualitative agreement with observations, e.g., \cite{aird_x-rays_2018}.

\begingroup
\renewcommand{\arraystretch}{1.5}
\begin{table}
    \centering
    \begin{tabular}{lcccc}
\hline
ID & \makecell{$v_{\mathrm{flow}}$ \\ (km\,s$^{-1}$)} & X--ray detection & WHaN & Ref. \\
\hline
\hline

34495 & 
-- &
Yes &
AGN &
1 \\

39063 &
-- &
No &
Weak AGN &
1 \\

45981 &
\hphantom{0}$-$550$\genfrac{}{}{0pt}{}{+120}{-80}$ & 
No &
Retired &
1 \\

50789 &
\hphantom{$-$}1290$\genfrac{}{}{0pt}{}{+320}{-220}$ &
No &
\makecell{Retired} &
1 \\

55410 & 
-- & 
No &
Retired &
1 \\

57000 &
\hphantom{$-$0}310$\genfrac{}{}{0pt}{}{+120}{-100}$ &
No &
AGN &
1 \\

65915 &
-- &
No &
Retired & 
1 \\

94982 &
-- &
No &
Retired &
2 \\

106260 &
-- &
Yes &
AGN &
1 \\

109760 & 
\hphantom{$-$0}470$\genfrac{}{}{0pt}{}{+80}{-100}$ &
No &
Weak AGN &
1 \\

113667 & 
-- &
No &
Weak AGN &
1 \\

117560 &
\hphantom{0}$-$320$\genfrac{}{}{0pt}{}{+170}{-140}$ &
No &
Retired &
1 \\

127460 &
-- &
No &
Weak AGN &
2 \\

\hline
\end{tabular}
    \caption{AGN properties of our sample. Column 5 notes the work from which the classification of each galaxy on the WHaN diagram was taken. Reference~1 is \protect\cite{stevenson_primer_2025} and reference 2 is \protect\cite{skarbinski_jwst_2025}.}
    \label{tab:agn}
\end{table}
\endgroup

\subsection{AGN Activity in our Sample}\label{sub:drivers}

We now discuss various indicators of AGN activity in our galaxies. We summarise the AGN classifications of our sample in Table \ref{tab:agn}. Overall, as discussed in Section \ref{section:energetics}, there seems to be no strong correspondence between current AGN activity (as diagnosed by X--ray detections or optical emission lines) and outflow incidence.

\subsubsection{Optical Emission Lines}\label{ref:oiii}

In a companion paper, \cite{stevenson_primer_2025} places the $z \geq 3$ objects in our sample onto the WHaN diagram \citep[][]{cid_fernandes_comprehensive_2011}. The WHaN diagram is commonly used to differentiate between AGN activity, star-formation and ionization from hot low-mass evolved stars (HOLMES), via comparison of H$\alpha$ equivalent widths and [N\,\textsc{ii}]/H$\alpha$ line ratios. The two remaining objects in our sample (EXCELS--94982 and EXCELS--127460) are included in a similar analysis for $z < 3$ galaxies in \cite{skarbinski_jwst_2025}.

From Table \ref{tab:agn} it can be seen that 7/13 of our galaxies are classified as AGN or weak AGN on the WHaN diagram. Based upon our \textsc{eagle} predictions (Fig. \ref{fig:eagle}) this level of AGN incidence would imply very low average Eddington ratios for these objects of log($\lambda)\simeq-5\pm1$. We can attempt to check that this is plausible by estimating Eddington ratios from optical line fluxes. However, even for objects with WHaN classifications of AGN, there is likely a significant contribution to H$\alpha$ from HOLMES. The high-ionization [O\,\textsc{iii}] 5007\AA\ line, however, should be a purer indicator of AGN activity.

We convert between [O\,\textsc{iii}] luminosities (or upper limits) for our galaxies and bolometric AGN luminosities using $L_{\mathrm{Bol,\, [O~\textsc{iii}]}} = 600\,L_{\mathrm{[O~\textsc{iii}]}}$, as in Section \ref{sub:agnwinds}. We can then divide this by the Eddington luminosity, $L_{\mathrm{Edd}}$, given by:
\begin{equation}
    L_{\mathrm{Edd}} = 1.26 \times 10^{38} \bigg(\frac{M_{\mathrm{BH}}} {{\mathrm{M}}_{\odot}}\bigg)\,\,\mathrm{erg\,s}^{-1}
\end{equation}
to obtain Eddington ratios. Here, $M_{\mathrm{BH}}$ is the black hole mass of the galaxy, which we approximate using the \cite{sun_no_2025} $M_{\mathrm{BH}}-M_{*}$ relation for galaxies at $1 < z < 4$, of $\mathrm{log_{10}}(M_{\mathrm{BH}}/M_{*}) = -2.5\pm0.3$.

For a galaxy with the average stellar mass for our sample, $\mathrm{log}_{10}(M_{*}/\mathrm{M}_{\odot}) = 10.75$, at $z=3$, an Eddington ratio of log($\lambda)=-5$ implies an [O\,\textsc{iii}] flux of $5\times10^{-21}$ erg s$^{-1}$ cm$^{-2}$, before any dust attenuation. The detection threshold for [O\,\textsc{iii}] in our spectra is $\sim10^{-19}$ erg s$^{-1}$ cm$^{-2}$, and only 2/13 objects have [O\,\textsc{iii}] detections above this level, both with WHaN classifications of `AGN'. This means that 5/7 objects classified as AGN or weak AGN do not have [O\,\textsc{iii}] detections, which is consistent with these objects having, on average, very low Eddington ratios, as implied by \textsc{eagle}. For the remaining two objects, we estimate Eddington ratios from their [O\,\textsc{iii}] fluxes of $\lambda=0.004\pm0.002$ for EXCELS--57000 and $\lambda=0.05\pm0.02$ for EXCELS--34495. This is consistent with the former not being X-ray detected, whilst the latter is (see Section \ref{ref:xray}).

As demonstrated in Section \ref{sub:agnwinds}, it is implausible that AGN activity at these very low levels could drive the observed outflows. For the purposes of our outflow cycle model therefore, we now turn to X--ray activity in our sample as a tracer of higher-Eddington-ratio AGN accretion.

\subsubsection{X--Ray Detections}\label{ref:xray}

Our sample benefits from deep $Chandra$ X--ray coverage in the $0.5-8$ keV band from the XUDS survey \citep{kocevski_x-uds_2018}. From our sample, 2/13 objects (EXCELS--34495 and EXCELS--106260) are X--ray detected, and neither have signs of outflowing neutral gas as indicated by their Na\,\textsc{d} absorption profiles. \cite{almaini_no_2025} investigated the prevalence of X--ray detected PSBs at $1 < z < 3$, finding a detection rate of $\sim$ 6 per cent. The authors additionally find weak AGN activity when stacking PSBs with no individual detections, concluding that PSBs typically show only weak levels of AGN activity at any given time, much weaker than in star-forming galaxies, and comparable to older passive galaxies. 

We can determine the Eddington ratio, $\lambda$, required to be X--ray detected at $z = 3$, via:
\begin{equation}
    \lambda = \frac{18\,L_{\mathrm{X, lim}}}{L_{\mathrm{Edd}}},
\end{equation}
where $L_{\mathrm{X, lim}}$ is the limiting X--ray luminosity required for detection, and 18 is the bolometric correction factor used by \cite{almaini_no_2025}, which assumes a photon index, $\Gamma=1.7$.

The median X--ray flux limit in XUDS is $8.4\times10^{-16}$\,erg\,s$^{-1}$\,cm$^{-2}$, corresponding to a limiting X--ray luminosity of $L_{\mathrm{X, lim}} = 10^{43.5}$\,erg\,s$^{-1}$. Following the assumptions in Section \ref{ref:oiii}, a $\mathrm{log}_{10}(M_{*}/\mathrm{M}_{\odot}) = 10.75$ galaxy at $z \sim 3$ would therefore require a black hole accreting at an Eddington ratio of $\lambda = 0.027$ to be detected in XUDS. For EXCELS--34495, the X-ray flux implies $\lambda=0.04\pm0.02$, in excellent agreement with the [O\,\textsc{iii}]-derived estimate of $\lambda=0.05\pm0.02$ from Section \ref{ref:oiii}. For EXCELS--106260, the X-ray flux again implies $\lambda=0.04\pm0.02$, however this object does not have detectable [O\,\textsc{iii}] emission. Given the [O\,\textsc{iii}] flux upper limit of $\sim 10^{-18}\,$erg\,s$^{-1}$\,cm$^{-2}$, approximately 2 magnitudes of dust attenuation on [O\,\textsc{iii}] would be required to explain this. There is some evidence from our spectral fitting that this object is among the dustiest in our sample (see also Leung et al. in prep).

We mark the Eddington ratio of $\lambda = 0.027$ required for X-ray detection with dashed magenta vertical lines in Fig. \ref{fig:eagle}. It can be seen from the top panel that \textsc{eagle} predicts $z=3$ quiescent galaxies spend $\simeq13$ per cent of their time above this Eddington ratio, in excellent agreement with the X--ray detection frequency of 2/13 in our sample. We therefore conclude that the AGN properties of our sample (as traced both by X--rays and rest-optical emission lines) are consistent with the predictions of \textsc{eagle}. This is also consistent with the results of \cite{pouliasis_active_2024}, who find that the X-ray luminosity function predicted by \textsc{eagle} is consistent with observations in our redshift range for galaxies with $\mathrm{log}_{10}(M_{*}/\mathrm{M}_{\odot}) > 10$ at all but the highest X-ray luminosities ($L_{\mathrm{X}} > 10^{44}$\,erg\,s$^{-1}$).

Based on these arguments and our findings in Section \ref{sub:agnwinds}, we adopt the ansatz that X--ray detection can be used as a proxy for AGN activity in our sample at a level that could plausibly drive the outflows we observe. Using Equation \ref{eq:agne}, the Eddington ratio of $\lambda=0.027 $ required for X-ray detection corresponds to $\dot{E}_\mathrm{AGN}\simeq10^{43.5}$ erg s$^{-1}$, which is within $1\sigma$ of the most energetic outflow we observe in our sample (see Table \ref{tab:disc}).

\subsection{Incidence of Outflows, Inflows and AGN Within our Sample}\label{sub:prob}

To estimate timescales for our simple `outflow cycle' model, we need to know not only the probability of observing an X--ray detected galaxy in our sample, $\mathrm{P}(\mathrm{X})$, a galaxy with an outflow, $\mathrm{P}(\mathrm{O})$, or a galaxy with an inflow, $\mathrm{P}(\mathrm{I})$, but also the probability of observing a galaxy with both an X--ray AGN and an outflow, $\mathrm{P}(\mathrm{X}\cap\mathrm{O})$, or a galaxy with both an X--ray AGN and an inflow, $P(\mathrm{X}\cap\mathrm{I})$. As can be seen from Table \ref{tab:agn}, for our observed sample of 13 galaxies, $N_{\mathrm{X}} = 2$, $N_{\mathrm{O}} = 3$, $N_{\mathrm{I}} = 2$,  $N_{\mathrm{X+O}} = 0$ and $N_{\mathrm{X+I}} = 0$. 

To estimate these 5 probabilities, we use a simulation-based inference methodology similar to the one outlined in \cite{stevenson_primer_2025}, which we briefly summarise here. We draw 50 million samples from prior distributions of $\mathrm{P}(\mathrm{X})$ and $\mathrm{P}(\mathrm{O})$, allowed to vary between 0 and 1. The prior distribution for $\mathrm{P}(\mathrm{X}\cap\mathrm{O})$ is allowed to vary between [$\mathrm{max}\{0,\, \mathrm{P}(\mathrm{X})+\mathrm{P}({\mathrm{O}})-1\}$, $\mathrm{min}\{\mathrm{P}(\mathrm{X}),\, \mathrm{P}({\mathrm{O}})\}$]. We then simulate a population of 13 objects for each draw, with a random assignment of AGN and outflows based on the drawn probabilities. We keep only populations for which $N_{\mathrm{X}}$, $N_{\mathrm{O}}$ and $N_{\mathrm{X+O}}$ match those observed in our sample. From the retained simulated populations, we find $\mathrm{P}(\mathrm{X}) = 0.19 \pm 0.10$, $\mathrm{P}(\mathrm{O})= 0.28 \pm 0.12$ and $\mathrm{P}(\mathrm{X}\cap\mathrm{O}) = 0.03 \pm 0.05$. These values represent posterior probability distributions for $\mathrm{P}(\mathrm{X})$, $\mathrm{P}(\mathrm{O})$ and $\mathrm{P}(\mathrm{X}\cap\mathrm{O})$ in our observed sample. We repeat this process for combinations of X-ray detections and inflows, additionally obtaining P(I) = 0.19$\pm$0.10, P(X+I) = 0.03$\pm$0.04.

\subsection{A Proposed Outflow Timeline}\label{sub:timeline}

Using the groundwork laid out in the previous sub-sections, we now turn to inferring a possible timeline, or cycle, for outflows in high-redshift recently quenched galaxies (our model is shown in Fig. \ref{fig:cycle}). As discussed in Section \ref{ref:xray}, we find that the average fraction of the time the \textsc{eagle} galaxies have spent above our $\lambda = 0.027$ AGN activity threshold (above which we assume the AGN is X--ray detectable and capable of driving outflows) post quenching is 0.13, or $\sim 130$ Myr per billion years. As shown in Fig. \ref{fig:eagle}, the average number of episodes above $\lambda = 0.027$ is $N_{\lambda = 0.027} = 22.4$ Gyr$^{-1}$. Thus, we predict the average galaxy in our sample will be visible as an X--ray AGN for $t_{\mathrm{AGN}} \simeq t_{\lambda = 0.027}/N_{\lambda = 0.027} \simeq 5\,$Myr at a time, roughly once every $t_{\mathrm{cycle}} \simeq 1/N_{\lambda = 0.027}\simeq40$ Myr.

In Section \ref{sub:prob}, we found the probability of a galaxy in our sample hosting an outflow is $\mathrm{P}(\mathrm{O})= 0.28 \pm 0.12$, which corresponds to an outflow visibility time within the cycle of 
\begin{equation}
t_{\mathrm{outflow}} =  t_{\mathrm{cycle}}\times\mathrm{P}(\mathrm{O})\approx 10 \pm 5\,\mathrm{Myr.}
\end{equation}

To assess the plausibility of this timescale, we take the average outflow velocity from our sample of $\sim 500$\,km\,s$^{-1}$ and multiply by $t_{\mathrm{outflow}}$ to obtain a simple estimate of the maximum distance travelled by the outflow during this time of $\sim 2-8$\,kpc. This is around five times the average $r_{\mathrm{eff}}$ of our galaxies, although here we have made the simplifying assumption of a constant outflow velocity. In the small number of studies that have spatially resolved sodium winds in massive, $z = 3$ galaxies, outflow extents of $\sim 1-3$\,kpc are found \citep[][]{deugenio_fast-rotator_2024, perez-gonzalez_accelerated_2025}. Outflows traced by sodium absorption have been found to extend up to $\sim 15$\,kpc in local galaxies \citep[e.g.,][]{rupke_quasar-mode_2017}. We therefore conclude that an outflow timescale of this order of magnitude seems plausible.

In Section \ref{sub:prob}, we found that the probability of observing a galaxy in our sample that hosts an X--ray AGN and an outflow simultaneously is $\mathrm{P}(\mathrm{X}\cap\mathrm{O}) = 0.03 \pm 0.05$. This corresponds to a time-scale of $t_{\mathrm{both}} =t_{\mathrm{cycle}}\times\mathrm{P}(\mathrm{X}\cap\mathrm{O})\approx 1 \pm 1$\,Myr, meaning that in our simple model outflows persist for $\simeq 9\pm5$\,Myr after an X--ray AGN has switched off. This agrees very well with the numerical modelling of \cite{zubovas_determining_2022}, which found that AGN (with $\lambda \geq 0.01$) should be visible for 6\,Myr and fossil outflows for 4\,Myr, when assuming a black hole growth phase of 10\,Myr. This is also very similar to the timeline proposed in \cite{almaini_no_2025} for recently quenched galaxies at cosmic noon (see their section 7.2).

Considering the above ($t_{\mathrm{AGN}} \sim 5$\,Myr, with outflows persisting for an additional 10\,Myr), we then have $\sim 25$\,Myr of the cycle remaining. 
After the AGN switches off, the outflow eventually stalls and dissipates, followed by a possible re-accretion of gas from the halo in the form of inflows. Two galaxies in our sample show evidence for inflowing gas. We can follow the same methodology as above to infer $t_\mathrm{inflow}=t_\mathrm{cycle} \times P(I) \simeq8\pm4$ Myr. 

With the simplifying assumption of an infall velocity matching that of the outflows ($\sim 500$\,km\,s$^{-1}$), and assuming the gas originates from a radius of $\sim 5$\,kpc, we approximate the gas inflow time as $t_{\mathrm{inflow}} \sim 10$\,Myr, again consistent with our cycle model. The inflowing gas may fuel further luminous AGN episodes, which in turn drive subsequent outflows, and the cycle repeats. Using the same arguments as above, the potential overlap between the end of the inflow phase and the start of the next inflow phase is again $\simeq1\pm1$ Myr. 

As a further consistency check on our inflow timescales, we can consider the free-fall time, which is of order
\begin{equation}
t_\mathrm{ff} \simeq \dfrac{1}{\sqrt{G\rho}}
\end{equation}
where G is the gravitational constant, and $\rho$ the mean density (e.g., \citealt{kippenhahn_stellar_1994}). For a massive quiescent galaxy with typical properties for our sample (we assume a stellar mass of log$_{10}(M_*/\mathrm{M_\odot})=10.75$ and a radius of 1 kpc) we obtain $t_\mathrm{ff}\sim4$ Myr. This is of the same order of magnitude as our inflow timescale estimate, and again demonstrates that this is physically plausible.

In summary, our observations appear to be consistent with quiescent galaxies at $z\simeq3$ undergoing a cycle with a duration of $\simeq40$ Myr on average, consisting of AGN activity ($\simeq5$ Myr), followed by outflows ($\simeq5-10$ Myr), then a lapse  ($\simeq20$ Myr), finally followed by an inflow phase ($\simeq5-10$ Myr) that reignites AGN activity. An indicative schematic for our model is shown in Fig. \ref{fig:cycle}. 
If some gas remains in the halo after each outflow, or escapes completely, the subsequent cycles will launch winds containing a lower gas mass, and eventually the galaxy will be entirely depleted of the cool, neutral gas needed to fuel further episodes of AGN activity (see also the discussion of escape velocities in Section \ref{sub:mout}). This is somewhat consistent with the results of \cite{taylor_high-velocity_2024}, who found that outflows persisted in quenched galaxies for only up to 1\,Gyr after their final starburst.

As discussed in Section \ref{section:intro}, post-starburst galaxies have been found to contribute an increasing fraction to the overall passive galaxy population with increasing lookback time \citep[e.g.,][]{wild_evolution_2016, taylor_role_2023}. At the redshifts probed in this work, it is likely that a large majority of quiescent galaxies will still be in (or have very recently left) a PSB phase, owing to the young age of the Universe \citep[e.g.,][]{deugenio_typical_2020}. To determine how representative of high-redshift quiescent galaxies our sample is, and to reveal the true impact of galactic winds at these epochs, larger samples of medium/high-resolution spectroscopy are crucial.

\section{Conclusions}\label{section:conc}

In this work, we present a sample of 13 post-starburst and quiescent galaxies at 1.8 $\leq z \leq$ 4.6, selected from the \textit{JWST} EXCELS survey. We utilise Na$\,$\textsc{d} profiles to investigate the incidence of outflowing and inflowing winds in these galaxies, and compare our sample to similar objects from recent literature. 

In Section \ref{sec:gf}, we classify our sample based on our best-fitting Na$\,$\textsc{d} models, finding outflow and inflow detection fractions of $23^{+13}_{-10}$ and $15^{+12}_{-7}$ per cent, respectively (see Figures \ref{fig:nadflow} and \ref{fig:nadnoflow}). These results are consistent with recent $z \sim 2$ findings (\citetalias{davies_jwst_2024}), along with studies of galaxies in the local Universe. We find outflow velocities of $300 \lesssim v_{\mathrm{flow}} \lesssim 1200\,\mathrm{km\,s}^{-1}$, and inflow velocities of $-550 \lesssim v_{\mathrm{flow}} \lesssim -300\,\mathrm{km\,s}^{-1}$. The remainder of our sample show no evidence for excess sodium absorption beyond the stellar continuum, agreeing with works at low redshift, although differing from detection rates for excess absorption at systemic velocity at cosmic noon (e.g., \citetalias{davies_jwst_2024}). The small sample sizes available at higher redshift make it difficult to draw firm conclusions on the cause of this. We investigate the possible effect of interactions and ongoing mergers on our sample in Section \ref{subsub:mergers}, and find that the likelihood of this having a significant impact on our outflow analysis is extremely low.

We compare wind velocity with time since quenching, $t_{\mathrm{sq}}$, in Section \ref{subsub:tsq}, finding that galaxies that have quenched within the last $\sim600$\,Myr are more likely to host an outflow (or inflow), for both EXCELS and our literature comparison sample (Fig. \ref{fig:tsq}). This is consistent with the results of \cite{taylor_high-velocity_2024}, who found that outflows in cosmic noon PSBs persist for up to 1 Gyr after a last starburst.

In Section \ref{section:energetics}, we compute mass, energy, and momentum rates for the outflows we find in EXCELS galaxies. We find mass outflow rates over two orders of magnitude higher than current levels of star formation in our sample, indicating that the observed winds may play a significant role in keeping the galaxy quenched. Comparing $v_{\mathrm{flow}}$ to $v_{\mathrm{esc}}$, we find two of the three galaxies in our sample with outflows have $v_{\mathrm{flow}}<v_{\mathrm{esc}}$, suggesting gas removed by the outflow may eventually fall back onto the galaxy, however the fate of the gas is hard to constrain. The remaining galaxy has $v_{\mathrm{flow}}\approx1.5v_{\mathrm{esc}}$, indicating that at least a fraction of the gas could escape the galaxy halo, reducing the available gas supply for future star formation. As typical mass loading factors in star-forming galaxies are $\eta \lesssim 10$, it is unlikely that these outflows are driven by residual star formation. We then compare our model-derived outflow energetics to expected energy and momentum rates from supernovae and AGN driven winds (Sections \ref{sub:sfwinds} and \ref{sub:agnwinds}, respectively). The energy and momentum outflow rates expected from supernovae are also insufficient to explain those observed for outflows in our sample (Fig. \ref{fig:sf}). 

Turning to AGN driven winds, we find one galaxy has outflow energetics that could potentially be explained by current AGN activity, while the others have anomalously high energy and/or momentum outflow rates compared to theoretical predictions (Fig. \ref{fig:agn}). We conclude that current levels of star formation and AGN activity could not plausibly drive the majority of the winds we observe in our sample, and thus we are likely observing fossil outflows driven by previous, more luminous AGN activity, which has since faded.

Motivated by this, in Section \ref{sec:time}, we construct a simple `outflow cycle' model (see Fig. \ref{fig:cycle}) based on the observed properties of our sample and guided by results from the \textsc{eagle} simulation. We conclude that our observations are consistent with a model in which quiescent galaxies at $z\simeq3$ undergo $\simeq5$ Myr periods of AGN activity strong enough to drive outflows every $\simeq40$ Myr on average. This AGN activity drives observable outflows that last for $\simeq10\pm5$ Myr, overlapping with the X-ray luminous AGN phase by $\simeq1\pm1$ Myr. This is followed by a $\simeq20$ Myr lull, then a $\simeq8\pm4$ Myr period of visible inflow, which eventually re-ignites AGN activity and the cycle repeats.

The large uncertainties in our inferred frequencies for AGN, inflows and outflows, and their resulting timescales, as well as the lack of correlation between outflow incidence and host galaxy properties are likely, in part, due to the small sample size studied here. This highlights the need for larger samples of high-resolution and high SNR spectroscopy of post-starburst and recently quenched galaxies at cosmic noon and beyond, as well as spatially resolved data, to help tackle uncertainties related to the geometry of the winds. Additional multiphase observations are crucial to pin down the molecular and ionized gas contents of the outflows. Finally, comparison across redshift is also a challenge, as different tracers of cool, neutral winds are accessible at different epochs. As well as the ever expanding JWST archive, upcoming instruments, such as the Multi Object Optical and Near-infrared Spectrograph for the VLT \citep[MOONS;][]{cirasuolo_moons_2020, maiolino_moonrise_2020}, will provide excellent spectroscopic samples at intermediate redshifts, including access to the crucial Na$\,$\textsc{d} feature.

\section*{Acknowledgements}
We thank the referee for their comments, which have improved the clarity of this manuscript. ET, ACC, H-HL and SDS acknowledge support from a UKRI Frontier Research Guarantee Grant (PI Carnall; grant reference EP/Y037065/1). OA acknowledges the support from STFC grant ST/X006581/1. FC, KZA-C, DS and TMS acknowledge support from a UKRI Frontier Research Guarantee Grant (PI Cullen; grant reference EP/X021025/1). VW acknowledges the Science and Technologies Facilities Council (ST/Y00275X/1) and Leverhulme Research Fellowship (RF-2024-589/4). JSD and DJM acknowledge the support of the Royal Society through the award of a Royal Society Research Professorship to JSD. Support for Program number JWST-GO-03543.014 was provided through a grant from the STScI under NASA contract NAS5-03127.

\section*{Data Availability}

All \textit{JWST} data products are available via the Mikulski
Archive for Space Telescopes (\url{https://mast.stsci.edu}).


\bibliographystyle{mnras}
\bibliography{refs} 




\appendix

\section{Comparison of Literature Samples}\label{appendix}

\begin{figure}
    \centering
    \includegraphics[width=\columnwidth]{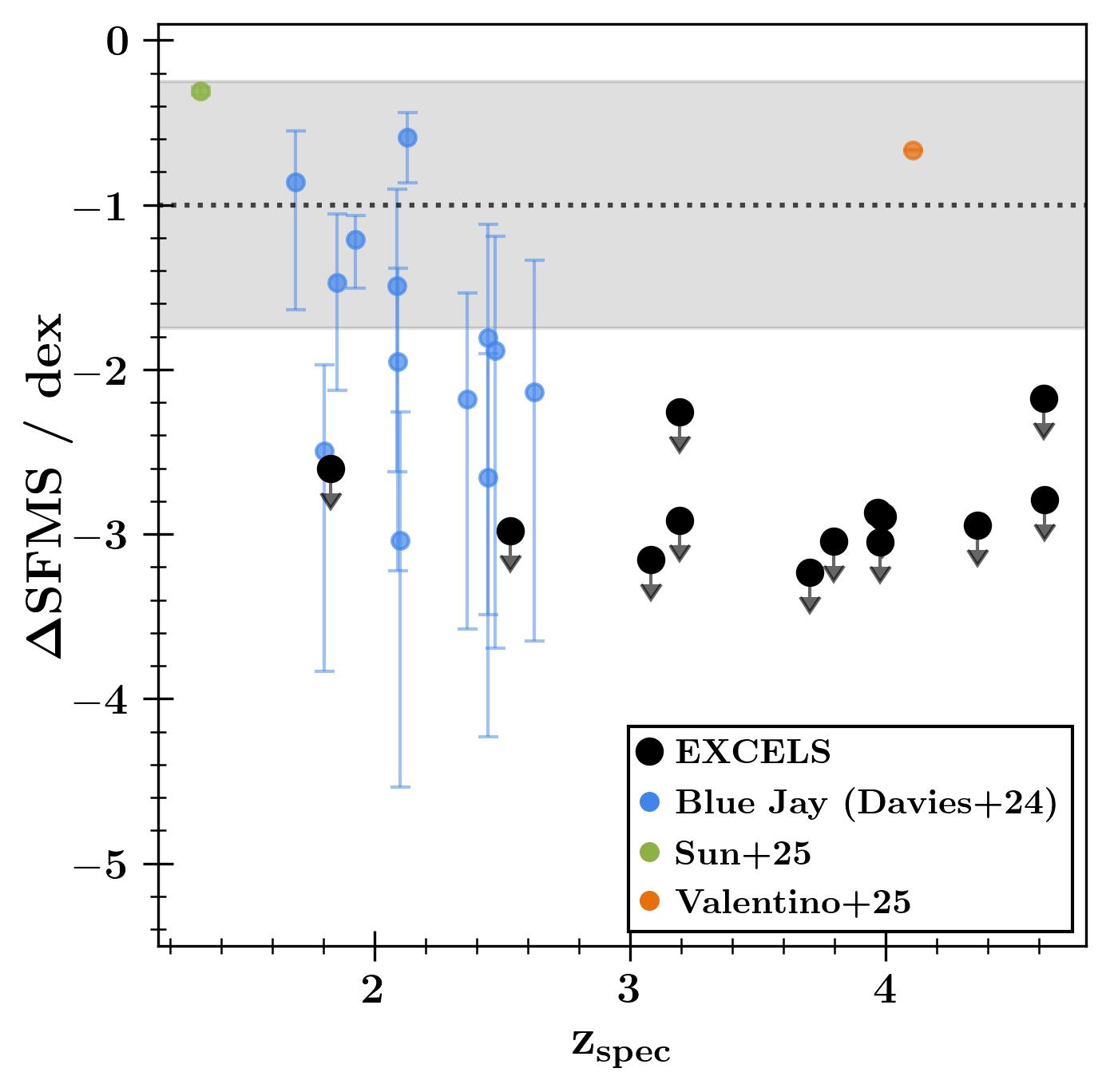}
    \caption{Comparison of the SFMS \citep[][]{leja_new_2022} selection method for the galaxies in our full sample. The grey shaded area represents the predicted error on the SFMS. Any galaxy below the dashed line is certain to have been selected as quiescent using this method.}
    \label{fig:sfms}
\end{figure}

\begin{figure}
    \centering
    \includegraphics[width=\columnwidth]{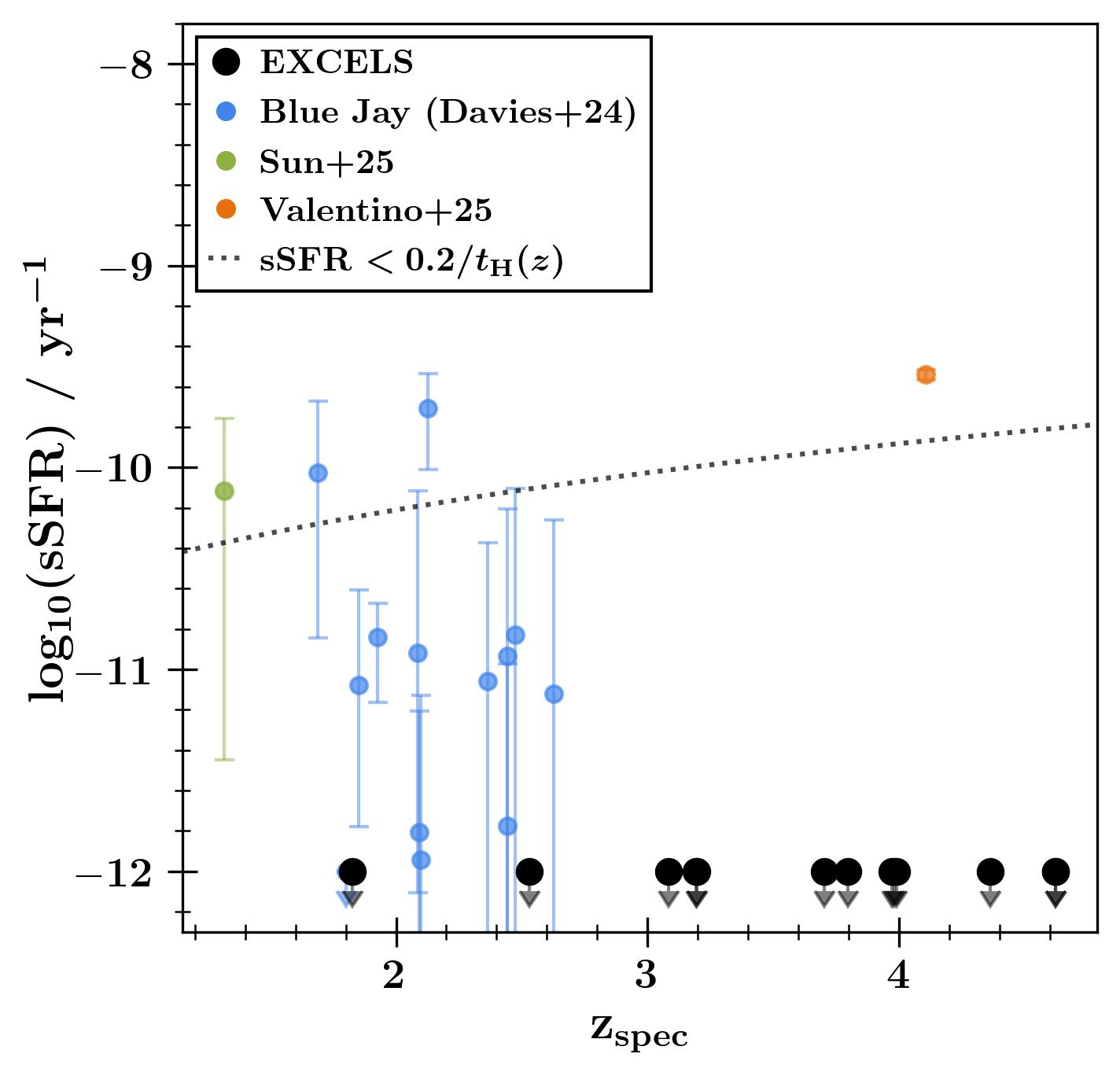}
    \caption{Comparison of the sSFR selection method for the galaxies in our full sample. Any galaxy below the dashed line is certain to have been selected as quiescent using this method.}
    \label{fig:ssfr}
\end{figure}

The additional literature comparison objects we consider in this work are outlined in Section \ref{sec:compsamp}. Here, we provide more detail in how the samples were selected, and compare these selection techniques to our own.

The Blue Jay survey \citep[\citetalias{davies_jwst_2024};][]{park_widespread_2024} selected quiescent galaxies as those 1\,dex or more below the star-forming main sequence (SFMS) at the median redshift of their sample, using the prescription of \cite{leja_new_2022}. In Fig. \ref{fig:sfms}, we show the offset from the SFMS for the EXCELS and comparison galaxies, where we have computed the SFMS at the observed redshift for each galaxy. The grey shaded region is the estimated error on the SFMS from \cite{leja_new_2022}. The full sample of EXCELS galaxies used in this work would be selected using this criterion.

Another commonly used quiescent galaxy selection method is a sSFR cut of sSFR < 0.2\,/\,$t_{\mathrm{H}}(z)$ (see Section \ref{sub:EXCELS}). In Fig. \ref{fig:ssfr}, we show this relation as a dashed line, and find, again, all EXCELS galaxies used in this work would be selected with this method. 

We additionally perform a quick test of whether the Blue Jay QG sample would be selected as PSBs or quiescent galaxies using our sample selection. We apply our selection method (outlined in Section \ref{sub:sample}) to the publicly available Blue Jay spectra\footnote{\href{https://jwst-bluejay.github.io/}{https://jwst-bluejay.github.io/}}. We find that 7/13 Blue Jay QG galaxies would be selected as PSBs, 1/13 would be selected as a quiescent galaxy, 2/13 would be discarded due to W$_{\textrm{[O$\,$\textsc{ii}]}}$ < $-5$\,\AA, and 3/13 would be discarded due to lack of H$\delta$ coverage. Overall, we would select 8 of the 10 Blue Jay QG galaxies with full coverage of the Na\,\textsc{d} and H$\delta$ features to be included in our sample, with the remaining 2/10 classified as star forming by our procedure.


\bsp	
\label{lastpage}
\end{document}